\documentclass[11pt]{article}
\usepackage{hyperref}
\usepackage{enumitem}
\usepackage{amsmath, mathrsfs, booktabs, amssymb,amsfonts,amsthm}
\usepackage{mathtools}
\usepackage{xcolor}
\usepackage{nicefrac}
\usepackage[round]{natbib}

\usepackage{bbold}

\protected\def\[#1\]{\begin{equation}\begin{aligned}#1\end{aligned}\end{equation}}
\protected\def\(#1\){\begin{equation*}\begin{aligned}#1\end{aligned}\end{equation*}}

\newcommand{\be}{\begin{equation*}\begin{aligned} }
\newcommand{\ee}{\end{aligned}\end{equation*} }

\newcommand{\bel}{\begin{equation}\begin{aligned} }
\newcommand{\eel}{\end{aligned}\end{equation} }

\newtheorem{theorem}{Theorem}
\newtheorem{lemma}{Lemma}
\newtheorem{corollary}{Corollary}

\theoremstyle{definition}
\newtheorem{example}{Example}

\theoremstyle{remark}
\newtheorem{remark}{Remark}

\renewenvironment{proof}[1][\proofname]{\bigskip\noindent {\bfseries Proof #1.}}{\qed}

\usepackage{graphicx}
\usepackage{color}

\usepackage{subcaption}
\usepackage{float}
\usepackage{comment}
\usepackage[capitalise,noabbrev]{cleveref}

\def\spacingset#1{\renewcommand{\baselinestretch}%
{#1}\small\normalsize}
\spacingset{1.2}

\begin{document}

\title{Bridged Posterior: Optimization, Profile Likelihood\\
	and a New Approach to Generalized Bayes}

\author{Cheng Zeng 
	\thanks{Joint first authors}
	\thanks{Department of Statistics, University of Florida, U.S.A. {czeng1@ufl.edu}}\quad\quad
	Eleni Dilma
	\footnotemark[1]
	\thanks{Department of Statistics, University of Florida, U.S.A. {edilma@ufl.edu}}\quad\quad
	Jason Xu
	\thanks{Department of Biostatistics, University of California Los Angeles, U.S.A. {jqxu@g.ucla.edu}}\quad\quad
	Leo L Duan
	\thanks{Department of Statistics, University of Florida, U.S.A. \href{email:li.duan@ufl.edu}{li.duan@ufl.edu}}
	\thanks{Corresponding author.}
}

\maketitle

\begin{abstract}
    Optimization is widely used in statistics, and often efficiently delivers point estimates on useful spaces involving structural constraints or combinatorial structure. To quantify uncertainty, Gibbs posterior exponentiates the negative loss function to form a posterior density. Nevertheless, Gibbs posteriors are supported in high-dimensional spaces, and do not inherit the computational efficiency or constraint formulations from optimization. In this article, we explore a new generalized Bayes approach, viewing the likelihood as a function of data, parameters, and latent variables conditionally determined by an optimization sub-problem. Marginally, the latent variable given the data remains stochastic, and is characterized by its posterior distribution. This framework, coined bridged posterior, conforms to the Bayesian paradigm. Besides providing a novel generative model, we obtain a positively surprising theoretical finding that under mild conditions, the $\sqrt{n}$-adjusted posterior distribution of the parameters under our model converges to the same normal distribution as that of the canonical integrated posterior. Therefore, our result formally dispels a long-held belief that partial optimization of latent variables may lead to underestimation of parameter uncertainty. We demonstrate the practical advantages of our approach under several settings, including maximum-margin classification, latent normal models, and harmonization of multiple networks.
\end{abstract}


\section{Introduction}
The generalized Bayes approach is becoming increasingly popular due to its potential advantages in model simplicity and robustness. A generalized posterior can be specified based on partial information from the data, or via a loss function that characterizes an inferential summary of the data. This appeals when the likelihood is inaccessible or intractable; there is a well-established literature on partial information settings including methods based on composite likelihood \citep{lindsay1988composite,varin2011overview}, partial likelihood \citep{sinha2003bayesian,dunson2005approximate}, pairwise likelihood \citep{jensen1994asymptotic}, and others. Recently, there has been a burgeoning interest in loss-based Bayesian models, including works involving classification loss \citep{polson2011data} or distance-based losses  \citep{duan2021bayesian,rigon2020generalized,natarajan2024cohesion}. Loss-based generalized Bayes models typically use a probability distribution called the Gibbs posterior \citep{jiang2008gibbs}, taking  the form:
$
\Pi(\theta\mid  y) \propto \exp\{ - g(\theta, y) \},
$
where $g(\theta, y)$ is some loss function with $y$ the data and $\theta$ the parameter.

There is a vast generalized Bayes literature using the Gibbs posterior for explicit model weighting, with $g$ chosen according to utility functions such as predictive accuracy \citep{lavine2021adaptive,tallman2024bayesian}, scoring rule \citep{gneiting2007strictly,dawid2015bayesian}, fairness metrics \citep{chakraborty2024gibbs} or summary statistics-based divergence \citep{frazier2021robust,frazier2022bayesian}. Such an approach also lends itself to modular descriptions of data \citep{jacob2017}, and can guard against model misspecification \citep{nott2023bayesian}.  With connections to these methods, our focus is on the case when one wants to adopt a  loss $g$ from the optimization literature for statistical modeling, while needing to quantify uncertainty beyond point estimates. 

The point estimate $\hat \theta = \arg\min_\theta g(\theta, y)$ can often be efficiently computed using an iterative optimization algorithm, even under a wide range of constraints. For example, convex clustering and its variants \citep{tan2015statistical,chi2015splitting,chakraborty2023b} use $g(\theta,y) = (1/2)\sum_{i=1}^n \| y_i- \theta_i\|^2_2 + \lambda\sum_{(i,j):i<j} \|\theta_i-\theta_j\|_2$ for data $y_i\in\mathbb{R}^p$, location parameter $\theta_i\in \mathbb{R}^p$, and tuning constant $\lambda>0$. This can be understood as a relaxation of hierarchical clustering; in place of a combinatorial constraint, the penalty term encourages most of the $L_2$-norms $\|\hat \theta_i-\hat \theta_j\|_2$ to be zero, promoting cluster structure via a small number of unique $\hat\theta_i$ at the solution.  The estimate $\hat \theta$ can be obtained using convex, continuous optimization. 
A popular combinatorial alternative makes use of the $k$-means loss \citep{macqueen1967some} toward clustering,  $g\{(c_1,\ldots,c_n),  y\}= \sum_{k=1}^K\sum_{(i<j):c_i=c_j=k} \| y_i- y_j\|^2_2 / n_k$, with $c_i\in\{1,\ldots,K\}$  where the discrete cluster assignment label $c_i=k$ if $\theta_k$ is the nearest centroid to $y_i$, $n_k =\sum_{i}1(c_i=k)$, and $\hat \theta_i = \sum_{i:c_i=k} y_i / n_k$. Here too, iterative algorithms can improve performance and avoid local minima using continuous optimization techniques \citep{xu2019}.   Recently,  \cite{rigon2020generalized} form a Gibbs posterior using this loss toward quantifying the uncertainty of $c_i$, which shows robustness to the distributional asymmetry. Several recent works import ideas from optimization to account for constraints within a Bayesian framework \citep{duan2020,presman2023,zhou2024proximal}.

Under the exponential negative transformation, the Gibbs posterior distribution concentrates near the posterior mode. This induces variability around the point estimate and, in turn, enables uncertainty quantification. 
How to interpret this uncertainty is not immediately obvious, and one may question the authenticity of inferential procedures such as hypotheses tests or intervals based on such a posterior which may not derive from a generative likelihood model. There are several  works that provide  justification in the large $n$ regime. First, Gibbs posteriors admit a coherent update scheme for $\theta$ toward minimizing the expected loss $\int g(\theta, y) \mathcal F(\textup{d}  y; \theta)$, where $\mathcal F$ denotes the true data generating distribution \citep{bissiri2016general}. Second, if the Gibbs posterior density is proportional to a composite
likelihood, such as the conditional density under some insufficient statistic \citep{lewis2021bayesian} derived from a  full likelihood $F(y; \theta_0)$,
then the Gibbs posterior of $\theta$ concentrates toward $\theta_0$ and enjoys asymptotic normality under mild conditions \citep{miller2021asymptotic}. 

These methodological and theoretical breakthroughs lend a cautious optimism that loss functions from the machine learning and optimization literature have the potential to broaden the scope of Bayesian probabilistic  modeling  
\citep{khare2015convex,kim2020bayesian,ghosh2021strong,martin2022direct,syring2023gibbs,winter2023sequential}. At the same time, two pitfalls of Gibbs posteriors motivate this article. The first is computational: the Gibbs posterior is often supported on a high-dimensional space, and fails to reduce the computational burden that often plagues posterior sampling schemes such as Markov chain Monte Carlo (MCMC) in high-dimensional problems. There is a large literature characterizing the scaling limit of MCMC algorithms, which can lead to slow mixing of Markov chains as the dimension of $\theta$ increases \citep{roberts2001optimal,belloni2009,johndrow2019mcmc,yang2019optimal}. Meanwhile, many semi-parametric models feature a low-dimensional $\theta$ as well as a latent variable whose dimension grows with $n$. When closed-form marginals are not available, the necessity of sampling these latent variables can also lead to critically slow mixing. These issues have been observed in popular statistical methods such as latent normal models, and have motivated a large class of approximation methods \citep{rue2009approximate} as alternatives to MCMC. This bottleneck explains in part the lack of Gibbs posterior approaches in latent variable contexts.

The second methodological gap relates to the modeling front: continuity of the Gibbs posterior distribution often yields a mismatch to constraint conditions on $\hat\theta$ except on a set of measure zero. 
To illustrate, consider the  Bayesian lasso \citep{park2008bayesian}, which can be viewed as the Gibbs posterior using the lasso loss. Though this promotes a sparse estimate $\hat\theta$, under its posterior distribution $\theta$ is non-sparse almost everywhere. A similar problem arises in a Gibbs posterior approach to support vector machines. The maximum-margin hyperplane has zero posterior measure, which may partially explain why studies from this view have focused on point estimation \citep{polson2011data}, and motivates our approach in seeking a more natural quantification of the associated uncertainty. Beyond this incongruence between  $\hat\theta$ and the samples from $\Pi(\theta\mid y)$,   invariance to changes in $g(\theta,y)$ presents another consideration. To obtain an estimate $\hat\theta$ residing in a constrained or low-dimensional space, it is common practice in optimization to employ an alternative  $\tilde g(\theta,y)$ that has superior computational properties. For example, $\tilde g$ can be convex, unconstrained, or non-combinatorial,  under the condition that  ${\arg\min}_\theta \tilde g(\theta,y)={\arg\min}_\theta g(\theta,y)$---that is, the two distinct loss functions touch at the minima. This invariance at the optimum is routinely exploited in methods such as convex relaxation, variable splitting, proximal methods, and majorization-minimization \citep{polson2015proximal,zheng2020relax,landeros2023}. However, the Gibbs posterior does not enjoy such an invariance, as the distribution  $\Pi(\theta\mid y)$ changes whenever $g$ changes.

These issues lead us to take a marked departure from existing approaches. Rather than treating $\theta$ as a high-dimensional random variable, we model $\theta=(z, \lambda)$ with only $\lambda$ as a parameter with a corresponding prior distribution. The argument $z$ is instead treated as a latent variable that is deterministic conditional on $y$ and $\lambda$, though importantly it remains a stochastic quantity when conditioned on $y$ alone. As we will demonstrate in the article, this effectively reduces the dimension of $\theta$ to nearly that of $\lambda$, simultaneously addressing both issues surveyed above. Specifying $z$ as the solution of an optimization subproblem allows us to retain transparent constraint conditions such as low rank, low cardinality, or  combinatorial constraints.

It is natural to ask whether such an approach is consistent with Bayesian methodology, that there exists a valid generative model corresponding to a likelihood that depends only on $\lambda$. This article answers this question affirmatively. We begin with a set of profile likelihoods that partially maximize a joint likelihood $L(y;z,\lambda)$ over $z$, showing that each corresponds to another common likelihood where the data are modeled dependently. We then establish the theoretical result that under mild conditions, the $\sqrt{n}$-adjusted posterior distribution of the parameter under our framework converges asymptotically to the same normal limit as canonical posteriors marginalized over non-deterministic latent variables. This contribution is closely related to prior work by \cite{polson2016mixtures}, which discovers a hierarchical duality: the scale mixture of univariate exponential or location-scale mixture of normal is proportional to another (potentially intractable) density maximized over a univariate latent variable. This perspective inspires efficient new algorithms for producing point estimates. Despite some similarities in the univariate setting, our method applies generally to multivariate problems and to settings where the latent variables may exhibit dependence. In other related work, \cite{lee2005profile} interpret the profile likelihood as resulting from an empirical prior. A key difference is that our proposed framework $L(y;z,\lambda)$ can lead to a fully Bayesian method, where the latent variable $z$ characterizes the latent dependency among the data. The source code can be found on \href{https://github.com/Zeng-Cheng/bridged_posterior_code_for_paper}{\color{blue} https://github.com/Zeng-Cheng/bridged\_posterior\_code\_for\_paper}.

\section{Method}
\subsection{Augmented Likelihood with Conditional Optimization}
To provide background, we first review the canonical likelihood involving {\em latent variables}, which takes the form
 \[\label{eq:canonical}
 L(y; \lambda) = \int  L(y ,  \textup d z ; \lambda)= \int L (y \mid z, \lambda) \Pi_{\mathcal L}( \textup d  z; \lambda),
 \]
in which we refer to $\lambda \in \mathbb{R}^d$ as the parameter,  and $z\in \mathbb{R}^p$ as the latent variable. Here $\Pi_\mathcal L$ denotes the marginal latent variable distribution for $z$. Since $z$ could be associated with a continuous, discrete, or degenerate distribution, we use the integration with respect to a probability measure notation $\int f(z) \mu(  \textup d z)$, in which $z$ with distribution $z\sim \mu$ is the one that we integrate over.
The joint distribution $L(y,z; \lambda)$ is also known as an augmented likelihood \citep{tanner1987calculation,van2001}. Examples abound in statistics: for instance, augmented likelihoods are used in characterizing dependence among discrete $y$ via a correlated normal latent variable $z$  \citep{wolfinger1993covariance,rue2009approximate}, or model-based clustering on grouping data $y$ via a latent discrete label $z$ \citep{blei2003latent,fraley2002model}. We now consider a special case when given $y$ and $\lambda$:
\[\label{eq:eq_constraint}
(z\mid y,\lambda) =  \hat z (y,\lambda) := {\arg\min}_{\zeta}  g ( \zeta,y; \lambda) \text{ with probability 1}.
\]
If the $\arg\min$ is unique, then $z$ is a \textit{conditionally deterministic} latent variable, which we abbreviate CDLV. Otherwise, $z$ has a conditional distribution supported on the solution set $\{{\arg\min}_{\zeta}  g ( \zeta,y; \lambda) \}$.

For simplicity of exposition, from here we focus on the case where $z$ is the {\em unique} minimizer. This encompasses a large class of models and is satisfied whenever $g(\zeta, y;\lambda)$ is strictly convex in $\zeta$ for every $(y,\lambda)$. Though $z$ is \textit{conditionally} deterministic, it is important to note that when we do not condition on $y$, $z$ remains randomly distributed under $\Pi_{\mathcal L}(  z;\lambda)$. This suggests a generative view according to \eqref{eq:canonical}: we have
\[\label{eq:constrained_view}
& z\sim \Pi_{\mathcal L}(  z;\lambda);  \qquad
 y\mid z,\lambda \sim  L(y\in\mathcal{Y}_{\lambda,z}\mid z, \lambda), \\
& \text{ where } \mathcal{Y}_{\lambda,z} = \bigl\{y: {\min}_{\zeta} g (\zeta,y; \lambda) = g (z,y; \lambda) \bigr\}.
\]
That is, $y$ is generated under the constraint given by $z$. This formulation allows the latent $z$ to have varying dimension $p$ and $\Pi_{\mathcal L}$  according to the sample size $n$.
	 
For concreteness, we present two illustrative examples based on the profile likelihood. Profile likelihoods have frequentist origins, motivated by the convenience of testing or constructing confidence intervals for a parameter of interest $\lambda$ in the presence of other \textit{nuisance parameters} $\zeta$. There is a long-standing debate on whether using a profile likelihood leads to a coherent Bayes' procedure \citep{lee2005profile,cheng2009penalized,evans2016measuring,maclaren2018profile}. Using the above, we can now view the profile likelihood as a special case of \eqref{eq:eq_constraint}, taking $g(\zeta, y; \lambda)= -\log L(y,\zeta;\lambda)$.

\begin{example}
    Consider linear regression with $y\in\mathbb{R}^n$, $X\in\mathbb{R}^{n\times d}$, $\lambda\in \mathbb{R}^d$, $z>0$, $v>0$:
    \(
    & L(y,z;\lambda) \propto z^{-n/2}\exp\biggl( - \frac{\|y-X\lambda\|^2}{2 z}   \biggr) z^{-v/2-1}\exp\biggl( - \frac{v}{2 z} \biggr) ;
    \\
    & z = {\arg\min}_\zeta   \{-\log L(y,\zeta;\lambda)\}.
    \)
    The first line has the same form as a likelihood with normal errors, with the variance $z$  regularized by an Inverse-Gamma($v/2$, $v/2$). Instead of marginalizing out $z$, we maximize $\log L(y,\zeta;\lambda)$ over $\zeta$ to obtain  $z= (v+\|y-X\lambda\|^2)/(v+n+2)$. Therefore, we have
    \(
    & \Pi_{\mathcal L}(z;\lambda) \propto z^{-(n+v+2)/2}; \\
    & L(y\mid z, \lambda) \propto \exp\bigg( - \frac{\|y-X\lambda\|^2}{2 z} \bigg) \mathbb{1}\biggl\{ \sum_{i=1}^n \bigl(y_i- x_i^\top \lambda\bigr)^2=(v+n+2)z -v \biggr\}.
    \)
    In particular, the indicator above imposes a conditional constraint $\mathcal Y_{\lambda,z}$  on $y$ given $z$, which corresponds to a ball centered at $X\lambda$ with radius $\sqrt{(v+n+2)z -v}$. Upon substituting an expression in $y$ for $z$, we obtain a marginal density
    \(
    L (y;\lambda ) \propto  \biggl\{ 1+\frac{\|y-X\lambda\|^2}{ (v+2)  \frac{v}{v+2}}\biggr\}^{-(n+v+2)/2},
    \)
    which coincides with the likelihood $L(y;\lambda)$ under an $n$-variate $t$-distribution with $v+2$ degrees of freedom, center at $(X\lambda)$, and covariance $\{v/(v+2)\}I$.
\end{example}

\begin{example}    Consider a multivariate factor model with $y = C z + \epsilon \in \mathbb{R}^{\tilde p}$, $\epsilon\sim \text{N}(0, I\sigma^2)$, $C\in \mathbb{R}^{\tilde p\times p}$.  Here let $\tilde p \ge  p$, the matrix $C$ have rank $p$,    $\lambda=(G, \sigma^2)$, and $G$ positive definite:
    \(
    & L(y,z ;\lambda) \propto \exp\biggl( - \frac{\|y-C z\|^2}{2 \sigma^2} \biggr) \exp\biggl( -\frac{1}{2} z^{\top} G^{-1} z \biggr); 
    & z = {\arg\min}_\zeta   \{-\log L(y,\zeta;\lambda)\}.
    \)
    The first part has the same form as a likelihood with $z$ regularized by a multivariate normal distribution $\text{N}(0,G)$. Here, minimization yields $z = (C^\top C/\sigma^2 +  G^{-1})^{-1} C^\top y/ \sigma^2 = \{G - G C^\top (I\sigma^2+ C GC^\top )^{-1} C G \}C^\top y/ \sigma^2$. Therefore, 
    \(
    &\Pi_{\mathcal L}(z;\lambda) \propto  \exp\biggl[ -\frac{1}{2} z^{\top} \{GC^{\top}  (I\sigma^2+ C GC^\top )^{-1}CG\}^{-1} z \biggr]; \\
    & L(y\mid z, \lambda) \propto \exp\biggl( - \frac{\|y-Cz\|^2} {2 \sigma^2}\biggr) \mathbb{1}\bigl \{
 C^\top y- (C^\top C +  \sigma^2 G^{-1}) z =0
    \bigr \}.
    \)
The indicator again imposes a conditional constraint $\mathcal Y_{\lambda,z}$ on $y$ given $z$, in this case corresponding to an affine subspace of $\mathbb{R}^{\tilde p}$ with dimension $\tilde p-p$. The marginal of $y$ is
    \(
    L(y;\lambda) \propto   \exp\bigg \{ -\frac{1}{2} y^\top  \Bigl( I\sigma^2 + C G C^\top \Bigr)^{-1} y 
    \bigg\},
    \)
    corresponding to a multivariate normal  $\text{N}\bigl(0, I\sigma^2 + C G C^\top\bigr)$.
\end{example}

From the above two examples, we highlight two observations: (i) partial optimization still leads to a valid probability kernel $ L(y;\lambda)$  associated with a coherent generative model for $y$; (ii) fixing $z$ at the conditional optimum induces dependency among the elements in $y$ in $L(y;\lambda)$, via the constraint $\mathcal Y_{\lambda,z}$.

The profile likelihood-based models are an important sub-class that we will primarily focus on. Nevertheless, in general, the loss function $g$ does not have to be the negative likelihood, and $z$ does not have to be available in closed form. We can still specify the joint likelihood, by including an optimization problem in the equality constraint \eqref{eq:eq_constraint}. 

\begin{remark}
    Despite the connection with the canonical full likelihood, in which $\lambda$ would be marginalized, the bridged posterior should be interpreted as a distinct generative model. Specifically, based on \eqref{eq:constrained_view} under the bridged posterior, $y$ has a distribution supported on a constrained space $\mathcal{Y}_{\lambda,z}$ given $z$. In contrast, the canonical full likelihood typically does not feature such a conditional constraint.
\end{remark}

\subsection{Bridged Posterior Distributions and Posterior Propriety}

We now take a Bayesian approach by assigning a suitable prior distribution on $\lambda$. Denoting this prior by $\pi_0(\lambda)$, Bayes theorem provides the posterior
\[\label{eq:bridge_post}
     \Pi (\lambda \mid y) = \frac{ \int L(y, \textup{d} z; \lambda )\pi_0(\lambda)}{\int\int L( y, \textup{d}z; \lambda )\pi_0( \textup{d} \lambda)},
    \quad \text{subject to }   z= {\arg\min}_{\zeta}  g (\zeta,y; \lambda).
\]
When $z$ is the unique minimizer, we may remove the first integration from both the numerator and denominator, replacing $\textup{d}z$ by $z$.
The above distribution can be viewed as obeying an {\em  equality constraint}, which acts as a bridge between a probabilistic model and an optimization problem. Therefore, we refer to \eqref{eq:bridge_post} as a bridged posterior. To clarify, the above formulation encompasses the setting of  $\lambda =(\lambda_A,\lambda_B)$, where only $\lambda_A$ influences the minimization of $g$, and $\lambda_B$ corresponds to the other parameters.

In the canonical setting, when $L(y,\zeta;\lambda)$ is the complete density/mass function of $(y,\zeta)$---so that it contains all the normalization with respect to $\lambda$---the integral $\int L(y,\zeta;\lambda) d \zeta$ is guaranteed to be a complete density/mass function of $y$. Here, specifying a proper prior $\pi_0(\lambda)$ is sufficient to ensure propriety of $\Pi(\lambda\mid  y)$. This is not automatically the case in our setting because $L(y,z;\lambda)$ may miss some normalizing terms. For instance in the generative view, $L(y\in\mathcal{Y}_{\lambda,z}\mid z, \lambda)$ fails to include the normalizing term associated with the constraint space $\mathcal{Y}_{\lambda,z}$.

A challenge arises when checking integrability (such as when verifying posterior propriety), when $L(y,z;\lambda)$ is intractable due to the lack of a closed-form solution $z$. Generally speaking, mathematically verification of integrability may vary from case to case; we develop a useful strategy in the case when $L(y,z;\lambda)$ is a profile likelihood. Consider the following 
\[\label{eq:primal}
    L(y,z;\lambda)= \exp\{- h(y,\lambda)\} \exp\{- \min_{\zeta}  g ( \zeta,y; \lambda)\},
\]
where $z=\arg\min_{\zeta}  g ( \zeta,y; \lambda)$. 

For checking posterior propriety, a common approach begins with finding an envelope function $f$, with $f(\lambda)\ge \pi_0(\lambda) L(y,z;\lambda)$ for all $\lambda$ in the parameter space, and then establishes that $f$ is integrable under the choice of $\pi_0$. To find such an envelope function, duality is a useful technique. To provide relevant background, we refer to $\min_{\zeta}  g ( \zeta,y; \lambda)$ as the \textit{primal problem}, and $z$ as the primal solution. Associated with the primal problem is the dual optimization problem $\sup_{\alpha} g^\dagger(\alpha, y;\lambda)$, where $\alpha\in \mathbb{R}^{q}$ is the dual variable.  For example, the Fenchel dual for convex $g$ is based on the conjugate function $g^\dagger(\alpha, y;\lambda) := \sup_{\zeta}\{ \alpha^\top \zeta - g ( \zeta,y; \lambda)\}$, and the Lagrangian dual for $\alpha$ under constraints $\tilde c(\alpha)\le \vec 0 $, where $\tilde c(\alpha)\in \mathbb{R}^q$ and the inequality holds pointwise, is $g^\dagger(\alpha, y;\lambda) := \inf_{\zeta \in \mathbb{R}^p} g(\zeta, y;\lambda) + \alpha^{\top } \tilde  c(\zeta)$, where the dual variable $\alpha \ge \vec 0$. 

The dual function $g^\dagger(\alpha, y;\lambda)$ is particularly useful here because of the weak duality. That is, $\sup_\alpha g^\dagger (\alpha,y;\lambda) \le {\inf}_{\zeta}  g (\zeta,y; \lambda)$ holds for any $\lambda$ in the feasible region, hence providing convenience for finding the envelope function.  We now state the useful bound:
\begin{theorem}\label{thm:duality}
    For a likelihood \eqref{eq:primal}, consider  ${\inf}_{\zeta}  g (\zeta,y; \lambda)$ as the primal problem, and \\$\sup_\alpha g^\dagger (\alpha,y;\lambda)$ as the dual problem with $E$ the feasible region of $\alpha$.
    If there exists $\tilde \alpha\in E$ such that $\int \exp[- h(y,\lambda)] \exp[-  g^\dagger ( \tilde \alpha,y; \lambda)] \pi_0( \textup d \lambda) < \infty$, then  $\int \Pi( \textup{d}\lambda\mid  y) < \infty$. 
\end{theorem}
\begin{remark}
   This result leads to a very useful method for checking integrability---we do not have to solve for the optimal dual variable $\hat\alpha$ at which $\sup g^\dagger (\alpha,y;\lambda)$ is attained. Instead, we just need to find any $\tilde \alpha \in E$ that makes the product integrable. Moreover, the criteria for weak duality are straightforward to check: for Fenchel duals,  $g$ needs to be convex, and for Lagrangian duals, $g$ can be convex or non-convex. We now illustrate the application of the theorem via a working example.
\end{remark}

\begin{example}[Latent normal model and latent quadratic exponential  model]\label{example:latent}
     We modify the canonical latent normal model that uses a full likelihood:
    \[\label{eq:latent_normal}
    \tilde L(y, \zeta;\lambda) & \propto \exp\biggl\{ -\frac{1}{2} \zeta^\top Q^{-1}(\lambda;x) \zeta \biggr\} \prod_{i=1}^n v(y_i \mid \zeta_i),
    \]
    where $v$ is commonly a log-concave density of $y_i$ conditionally independent for $i=1,\ldots,n$, $Q(\lambda;x)$ is  parameterized by a covariance kernel such as $Q(\lambda;x)_{i,j}= \tau \exp(- \|x_i-x_j\|^2/2b)$ with $x_i \in \mathbb{R}^{\tilde d}$ the observed predictor/location, and parameter $\lambda=(\tau,b)\in \mathbb{R}^2$. In our example, we focus on binary $y_i$ from Bernoulli distribution under logistic link $v(y_i \mid \zeta_i) = \exp(y_i \zeta_i) / \{1+\exp(\zeta_i)\}$. 
    We now minimize $g (\zeta,y; \lambda)= -\log \tilde L(y, \zeta;\lambda)$ over $\zeta\in \mathbb{R}^n$ to induce a conditionally deterministic $z$, with profile likelihood:
    \[\label{eq:latent_quadratic}
    & L(y, z;\lambda)  \propto  \exp\biggl\{ -\frac{1}{2} z^\top Q^{-1}(\lambda;x) z \biggr\} \bigg\{\prod_{i=1}^n v(y_i \mid z_i) \bigg\},
         & z = {\arg\min}_\zeta   \{-\log L(y,\zeta;\lambda)\}. 
    \]
    
    {We first show how to find the dual function}. As $g (\zeta,y; \lambda)$ can be conveniently decomposed into the sum of a quadratic function and a convex function, this lends itself to variable splitting  using the constraint $u=\zeta$. With $\alpha\in \mathbb{R}^{n}$ the Lagrange multiplier, we obtain the Lagrangian dual 
    \(
        g^\dagger (\alpha,y;\lambda) = \inf_{\zeta,u} \frac{1}{2} \zeta^\top Q^{-1} \zeta + \alpha^\top (\zeta-u) + \sum_{i=1}^n \bigl\{ -y_i u_i  + \log(1+ \exp(u_i)) \bigr\},
    \)
    where we use $Q= Q(\lambda;x)$ to ease notation. This leads to
    \(
        \hat \zeta = -Q\alpha,  \qquad \hat u_i =  \log \frac{\alpha_i+y_i}{1-(\alpha_i+y_i)} \text{ for }i=1,\ldots,n,
    \)
    whenever $(\alpha+y)\in (0,1)^n$; otherwise the infimum is $-\infty$. We have the dual function:
\(\label{eq:dual_cost_lnm}
        & g^\dagger (\alpha,y;\lambda) = -\frac{1}{2} \alpha^\top Q \alpha   - \sum_{i=1}^n \biggl\{ (a_i+y_i) \log \frac{\alpha_i+y_i}{1-(\alpha_i+y_i)}  - \log \frac{1}{1-(\alpha_i+y_i)} \biggr\}, \\
        & \text{subject to } (\alpha+y)\in (0,1)^n.
  \)
    At a given $\alpha$ satisfying $(\alpha+y)\in (0,1)^n$,  we have
    \(
        \exp\{ - g^\dagger (\alpha,y;\lambda) \}  = 
        \exp\biggl\{ \frac{1}{2} \alpha^\top Q(\lambda;x) \alpha \biggr\} \prod_{i=1}^n
          \bigg[ {(\alpha_i+y_i)^{a_i+y_i}}{\{1-(\alpha_i+y_i)\}^{1-(a_i+y_i)}} \bigg].
    \)
    Due to some similarity between the above form and the quadratic exponential model in \cite{mccullagh1994exponential}, we refer to \eqref{eq:latent_quadratic} as a latent quadratic exponential model.
    
        {We now show how to apply Theorem~\ref{thm:duality} to verify the posterior propriety}.     
    We can see that the above is an integrable upper bound for $y$, since $y_i\in \{0,1\}$, and both $(\alpha_i+y_i)$  and $1-(\alpha_i+y_i)$ are  bounded above. To find an appropriate prior for $\lambda$, the second part does not involve $\lambda$ at any fixed $\alpha$. It suffices to find a prior such that for a feasible $\tilde \alpha$:
    \(
         \int  \exp \biggl\{ \frac{1}{2}  \tilde \alpha^\top Q(\lambda;x) \tilde\alpha \biggr\} \pi_0(\textup{d}\lambda) <\infty.
    \)
     Using $Q(\lambda;x)_{i,j}= \tau \exp(- \|x_i-x_j\|^2/2b)$, the matrix spectral norm $\| Q(\lambda;x)\|_2\le  n \tau$.
    As we may choose any feasible $\tilde \alpha$, we take  $\tilde\alpha_i = -(1/n) \mathbb{1}(y_i=1) + (1/n) \mathbb{1}(y_i=0)$. Since $\tilde \alpha^\top Q(\lambda;x) \tilde \alpha \le \|\tilde\alpha\|_2^2    \| Q(\lambda;x)\|_2= \tau$, it suffices to assign a half-normal prior for $\tau$ proportional to $\exp(-c_1\tau^2)$ with $c_1>0$ and with any proper prior on $b>0$.
\end{example}

\begin{remark}
	For the above example, strong duality holds: $\sup_\alpha g^\dagger (\alpha,y;\lambda) = \inf_\zeta g (\zeta,y;\lambda)$. Therefore, we can use the dual ascent algorithm to find $\hat \alpha = {\arg\max}_{\alpha: (\alpha+y)\in (0,1)^n} g^\dagger (\alpha,y;\lambda)$, and then set $z = -Q \hat\alpha$. Note that neither the dual function \eqref{eq:dual_cost_lnm} nor its gradient with respect to $\alpha$ requires the inversion $Q^{-1}$, an $O(n^3)$ operation, so that optimization can be carried out very efficiently. At the same time, $L(y,z;\lambda)$ can be evaluated quickly since $z^\top Q^{-1}z = \hat\alpha^\top  Q\hat\alpha$.
	In contrast, the latent normal model would involve matrix inversion and decomposition for sampling latent $\zeta$. We defer the numerical experiments to Section \ref{sec:numeric}.
\end{remark}

In general, the function $g$ does not have to be the negative log-likelihood. In the \cref{sec:flow}, we provide a high-dimensional example of maximal flow problem. The optimization subproblem corresponds to a mechanistic process where flows automatically fill the network to the capacity, whereas the likelihood characterizes the difference between the conditionally optimal flows and observed values.

\subsection{Predictive Distribution}

In addition to parameter estimation, one may be interested in making predictions on data $y_{(n+1):(n+k)}$ and quantifying their uncertainty, using the following distribution:
\(
\Pi\bigl\{ y_{(n+1):(n+k)} \mid y_{1:n}\bigr\} & \propto \int L\bigl\{y_{(n+1):(n+k)}\mid y_{1:n}, \lambda\bigr\} \Pi( \textup{d}\lambda \mid y_{1:n}) \\
 & \propto \int 
\frac{L\{ y_{1:(n+k)}, \hat z(y_{1:(n+k)},\lambda), \lambda\}}{L\{ y_{1:n}, \hat z(y_{1:n},\lambda), \lambda\}}
\Pi( \textup{d}\lambda \mid y_{1:n}),
\)
for which we could take each posterior sample of $\lambda$, and simulate a vector $y_{(n+1):(n+k)}$ with kernel proportional to ${L\{ y_{1:(n+k)}, \hat z(y_{1:(n+k)},\lambda), \lambda\}}$. {We use $\propto$ to mean that $\Pi\bigl\{ y_{(n+1):(n+k)} \mid y_{1:n}\bigr\}$ contains the additional normalizing term, so it does integrate to one over $y_{(n+1):(n+k)}$.}

When we lack a way to directly draw from the joint distribution of $y_{(n+1):(n+k)}$, note that
\(
\frac{L\{ y_{1:(n+k)}, \hat z(y_{1:(n+k)},\lambda), \lambda\}}{L\{ y_{1:n}, \hat z(y_{1:n},\lambda), \lambda\}}=
\prod_{j=1}^{k}\frac{{L\{ y_{1:(n+j)}, \hat z(y_{1:(n+j)},\lambda), \lambda\}}}{{L\{ y_{1:(n+j-1)}, \hat z(y_{1:(n+j-1)},\lambda), \lambda\}}},
\)
suggesting that we can simulate $y_{n+j}$ sequentially for $j=1\ldots k$. When $y_{n+j}$ lies in a low-dimensional (often one-dimensional) space, we can employ a simple algorithm such as rejection sampling.
Note that {\em all} elements in $z=\hat z(y_{1:(n+k)},\lambda) \in \mathbb{R}^p$ may vary according to $y_{1:(n+k)}$; we emphasize this by using the
notation $\hat z(y_{1:(n+k)},\lambda)$.
This implies that when there is no closed-form solution for $z$, there is an additional burden to compute $\hat z(y_{1:(n+j)},\lambda)$ wherever $j$ increments to $j+1$. Fortunately, for the  problem
$
\hat z(y_{1:(n+j+1)},\lambda)= {\arg\min}_{\zeta}  g (\zeta,y_{1:(n+j+1)}; \lambda),
$
we can initialize $\zeta$ at the last optimal when predicting $y_{n+j},$ $\hat z(y_{1:(n+j)},\lambda)$, and it takes a few iterations of optimization steps to converge to $\hat z(y_{1:(n+j+1)},\lambda)$. For advanced problems, there is a large literature on online optimization algorithms \citep{jadbabaie2015online} that can be employed to efficiently obtain sequential updates.

For concreteness, we highlight a useful property of the above predictive distribution in the context of classification problems. We can find a hyperplane that not only divides the fully observed data with both predictors $x_i$ and labels $y_i$ for $i=1,\ldots,n$, but also seeks to separate the observed ``unlabeled'' data $x_{i'}$ with corresponding (unobserved) label $y_{i'}$,  $i'=n+1,\ldots, n+k$. This further improves the classification accuracy, and is often called the ``semi-supervised setting'' in the machine learning literature \citep{chapelle2010semi}.

\begin{example}[Bayesian Maximum Margin Classifier for Partially Labeled Data]\label{example:svm}
	Consider the following likelihood that extends the support vector machine \citep{cortes1995support}, for $n$ labeled data $(x_i,y_i)\in \mathbb{R}^{\tilde p} \times \{-1,1\}$  and $k$ unlabeled predictors $x_j \in \mathbb{R}^{\tilde p}$:
	\[\label{eq:svm}
	& L \big[ \{y_i\}_{i=1}^{n}, z=(z_w,z_b);\lambda, \{x_i\}_{i=1}^{n+k}\big] \propto   
	\sum_{ \{y_{n+j}\}_{j=1}^k \in \{-1,1\}^k}\exp \bigg\{-  \frac{1}{2}\lambda \| z_w\|^2_2 -\sum_{i=1}^{n+k} h ( z,y_i; \lambda,x_i) \bigg\}, \\
	& \text{subject to } z = {\arg\min}_{\zeta = (\zeta_w,\zeta_b)} 
	\frac{1}{2}\lambda \| \zeta_w\|^2_2
	+ \sum_{i=1}^{n+k} h ( \zeta,y_i; \lambda, x_i),
	\\
	& h ( \zeta,y_i; \lambda,x_i)= \max \bigl\{ 1- y_i (\zeta_w ^{\top} x_i + \zeta_b),0\bigr\},
	\]
	where $z_w\in \mathbb{R}^{\tilde p},\zeta_w\in \mathbb{R}^{\tilde p}$ and $z_b\in \mathbb{R},\zeta_b\in \mathbb{R}$. We treat $x_i$ as fixed, so that the above likelihood is viewed as a discrete distribution for $(y_1,\ldots,y_{n+k})$. The function $h$ is the hinge loss, which takes value zero when $y_i=1$, $\zeta_w ^{\top} x_i + \zeta_b \ge 1$, or when $y_i= -1$, $\zeta_w ^{\top} x_i + \zeta_b \le -1$. Effectively, the loss function penalizes not only the misclassified points $(x_i,y_i): y_i (\zeta_w ^{\top} x_i + \zeta_b)<0$, but also the points in the {\em band} between two boundaries 
	$\{x:  -1< \zeta_w ^{\top} x_i + \zeta_b<1\}$. The inclusion of $(1/2)\lambda \| z_w\|^2_2$ leads to a maximum distance between the two hyperplanes $\{x: z_w^{\top} x + z_b=1\}$ and $\{x: z_w^{\top} x + z_b=-1\}$, under some tolerance to non-zero hinge losses, with tolerance controlled by $\lambda>0$.

	For comparison, if we were to directly use a Gibbs posterior with likelihood of the form of \eqref{eq:svm}---that is, without the equality constraint so that  $z$ is replaced by $\zeta$, then it would hold that 
	\(
	p(y_{i} \mid \zeta,\lambda,x_i)  \propto
	{
		\exp\{-h(\zeta,y_i;\lambda,x_i)\} 
	}
	\)
	independently for $i=n+1,\ldots, n+k$. I particular, the distribution under the Gibbs posterior would yield $\tilde L\big[ \{y_i\}_{i=1}^{n}, \zeta;\lambda, \{x_i\}_{i=1}^{n+k}\big]
	=\tilde L\big[ \{y_i\}_{i=1}^{n}, \zeta;\lambda, \{x_i\}_{i=1}^{n}\big]
	$ via marginalization, which fails to incorporate any information from the observed $(x_{n+1},\ldots,x_{n+k})$. See \cite{liang2007use} for a more comprehensive discussion on this issue.
	
	\begin{figure}[H]
		\centering
		\includegraphics[width=0.6\linewidth]{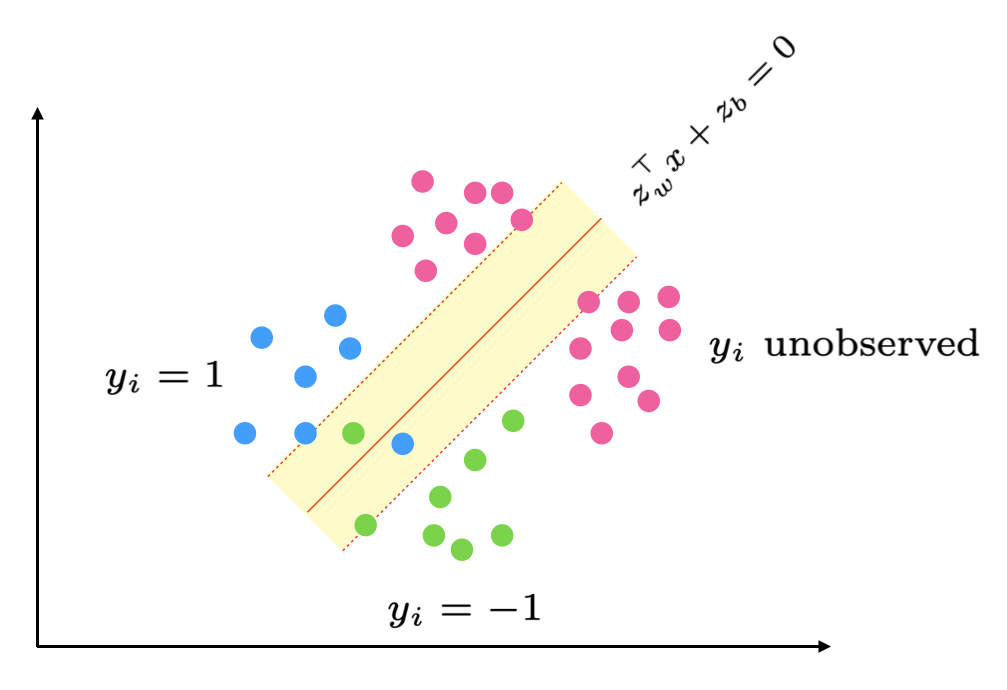}
		\caption{Intuition on how the Bayesian maximum margin classifier (a bridged posterior based on conditional minimization of the hinge loss) incorporates information from both the labeled ($y_i$ observed) and unlabeled data ($y_i$ unobserved). The posterior puts a high probability on a decision boundary with a small misclassification error among observed data (blue and green points), while trying to avoid having the decision band $\{x:-1< \zeta_w ^{\top} x + \zeta_b<1\}$ cover the unlabeled predictors $x_j$ (magenta points).
			\label{fig:sssvm_intuition}}
	\end{figure}
	
	Now, under our bridged posterior approach, denote the conditional optimum $z= \hat z\bigl(\{y_i\}_{i=1}^{n+k},\lambda; \{x_i\}_{i=1}^{n+k}\bigr)$. Though a closed-form marginal for \eqref{eq:svm} is not available, we know from the Lagrangian dual  with multiplier $\alpha\in \mathbb{R}^{n+k}$ \citep{chang2011libsvm} that the decision hyperplane $\mathcal C_z=\{x:z_w ^{\top} x + z_b=0\}$ satisfies
	\(
	z_w = \sum_{i=1}^{n+k} (\alpha_i y_i)  x_i, \quad \text{where} \quad 0\le \alpha_i\le \lambda^{-1},  \)
	and there are only a few $\alpha_i\neq 0$ for which $(\zeta_w ^{\top} x_i + \zeta_b)y_i\le 1$---these are  the so-called ``support vectors". That is, regardless of the values of
	$\{y_i\}_{i=n+1}^{n+k}$, the decision boundary can be influenced by the unlabeled predictor $\{x_i\}_{i=n+1}^{n+k}$. 
	Intuitively, the bridged posterior assigns higher probability to a hyperplane $\mathcal C_z$ that has a small misclassification error among observed data, while avoiding unlabeled predictors $x_i$ in the band $\{-1< \zeta_w ^{\top} x + \zeta_b<1\}$. We use Figure \ref{fig:sssvm_intuition} to illustrate the intuition.
\end{example}

\section{Posterior Computation}
	One appealing property of the bridged posterior is that the joint distribution $\Pi(\lambda,z \mid y)$ is supported on a low dimensional space relative to the ambient space, with intrinsic dimension determined by $\lambda$. This leads to efficient posterior estimation via MCMC algorithms.
	\subsection{Metropolis--Hastings with Conditional Optimization}
	We first focus on the case of $\lambda\in \mathbb{R}^d$ with a small $d$, which allows us to use simple MCMC algorithms such as Metropolis-Hastings for posterior sampling. At MCMC iteration $t$, we denote the posterior kernel as:
	$$ \Pi \bigl\{\lambda^{(t)} \mid y\bigr\} = m L\big\{ y, z^{(t)}; \lambda^{(t)} \bigr\} \pi_0(\lambda^{(t)}), \qquad z^{(t)}=  \hat z (y,\lambda^{(t)})= {\arg\min}_{\zeta}  g (\zeta,y; \lambda^{(t)} ).$$
	where $m$ is the normalizing constant that does not involve $\lambda^{(t)}$ or $z^{(t)}$. We assume that $\pi_0$ has a closed form and $L$ has a closed form as a function of $\bigl\{y^{(t)},z^{(t)},\lambda^{(t)}\bigr\}$, although $z^{(t)}$ may not have a closed form.  This allows us to use the following simple Metropolis--Hastings step:
	\begin{itemize}
		\item Draw proposal $\lambda^* \sim G(\cdot;\lambda^{(t)})$
		\item Run optimization subroutine to find $z^*=  {\arg\min}_{\zeta}  g (\zeta,y; \lambda^*).$
		\item Set $\lambda^{(t+1)} \leftarrow \lambda^*$, $z^{(t+1)} \leftarrow z^*$ with probability:
		\(
		1\wedge\frac{L( y, z^*; \lambda^* ) \pi_0(\lambda^*) G(\lambda^{(t)};\lambda^* )}{L\bigl\{ y, z^{(t)}; \lambda^{(t)} \bigr\} \pi_0(\lambda^{(t)}) G(\lambda^* ; \lambda^{(t)} )}.
		\)
		Otherwise, set $\lambda^{(t+1)} \leftarrow \lambda^{(t)}$, $z^{(t+1)} \leftarrow z^{(t)}$.
	\end{itemize}
	In this article, for algorithmic simplicity, we take $\lambda$ as unconstrained in $\mathbb{R}^d$ under appropriate reparametrization (such as the softplus transformation for positive scalars $\tilde\lambda_1 = \log[1+\exp(\lambda_1)]>0$).
	We use $G(\cdot;\lambda^{(t)})$ as $\text{Uniform}(\lambda^{(t)} - s ,\lambda^{(t)} + s)$, where $s\in\mathbb{R}^{d}_{\ge 0}$ is a tuning parameter that controls the step size in each dimension. 
	When running MCMC for each of the examples presented, we make use of an adaptation period to tune $s$ so that the empirical acceptance rate is close to $0.3$, after which we fix $s$ and collect Markov chain samples. This exhibits excellent mixing performance empirically.
	
	\subsection{Diffusion-based Algorithms for Profile Likelihood-based Bridged Posterior}

	Instead of uniform random walk proposals $G$, informative proposals such as the  Metropolis-adjusted Langevin algorithm (MALA) or Hamiltonian Monte Carlo may yield better performance. These algorithms become especially advantageous compared to random walk methods in terms of mixing as the dimension $d$ increases.
	
	In this subsection, we first show that gradients (or sub-gradients) are readily available in cases when a profile likelihood is used, and then we discuss its use in the MALA algorithm.
	Under the bridged posterior, the lack of closed forms for $z$ presents a potential challenge to these methods, leading to intractable gradients or subgradients with respect to $z$. However, for those based on the profile likelihood, this issue can be bypassed entirely. Consider the posterior deriving from \eqref{eq:primal},
	\(
	\Pi(\lambda \mid y) \propto    \pi_0(\lambda)\exp\{- h(y,\lambda)\} \exp\{- \min_{\zeta} g(\zeta,y;\lambda)\}, \quad   z= {\arg\min}_{\zeta}  g (\zeta,y; \lambda).
	\)
	If $g(\zeta,y;\lambda)$, as an unconstrained function of three inputs, is differentiable in $\zeta$ and $\lambda$ almost everywhere, then we have a very simple gradient expression provided $z= \arg\min_{\zeta} g(\zeta,y;\lambda)$ is  differentiable with respect to $\lambda$:
	\(
	\frac{ \partial \min_{\zeta} g(\zeta,y;\lambda)}{\partial \lambda} = \frac{\partial g(\zeta,y;\lambda)}{\partial \lambda} \bigg\vert_{\zeta=  z }.
	\)
	This is due to the envelope theorem.
	
	When $z$ may not be differentiable in $\lambda$ but is strictly continuous in  $\lambda$, the expression  ${\partial g(\zeta,y;\lambda)}/{\partial \lambda} \big\vert_{\zeta=  z }$ still holds as a  subgradient of $\min_{\zeta} g(\zeta,y;\lambda)$ with respect to $\lambda$ \citep[Theorem 10.49]{rockafellar2009variational}. For completeness, recall a subgradient
	of $f:\mathbb{R}^d\to \mathbb{R}$ at $x\in \mathbb{R}^d$ is a vector $v\in\mathbb{R}^d$ that satisfies  $f(y)\ge f(x)+ v^\top (y-x)$ for any $y$ in the domain. In subgradient-based MCMC samplers \citep{tang2024computational}, one typically refers to a local subgradient with inequality held for
	$y: \|y-x\|\le \epsilon$ under a sufficiently small $\epsilon>0$. When $f$ is differentiable at $x$, there is a unique subgradient, coinciding with the usual gradient.
	
	We use $\tilde\nabla \log \Pi(\lambda \mid y)$ to denote a subgradient evaluated at point $\lambda$. For reversibility, in the case when there is more than one subgradient at $\lambda$, we impose a constraint that $\tilde\nabla \log \Pi(\lambda \mid y)$ is chosen as one of the subgradients in a pre-determined way. This constraint is implicitly satisfied in most computing software, for example, most packages will output $\tilde\nabla |\lambda_1|_1=0$ when $\lambda_1=0$, even though any value $[-1,1]$ is a subgradient. We now describe the MALA algorithm with preconditioning.
	\begin{itemize}
		\item Draw proposal $\lambda^* \sim \text{N} \bigl[\cdot ;  \lambda^{(t)} + \tau M \tilde\nabla \log \Pi\{\lambda^{(t)} \mid y\}, 2\tau M \big].$
		\item Run optimization algorithm to find $z^*=  {\arg\min}_{\zeta}  g (\zeta,y; \lambda^*).$
		\item Set $\lambda^{(t+1)} \leftarrow \lambda^*$, $z^{(t+1)} \leftarrow z^*$ with probability:
		\(
		1\wedge\frac{L( y, z^*; \lambda^* ) \pi_0(\lambda^*) 
			\textup{N}\bigl \{\lambda^{(t)} ;   \lambda^* + \tau M \tilde\nabla \log \Pi(\lambda^* \mid y),  2\tau M \bigr\} 
		}{L\{ y, z^{(t)}; \lambda^{(t)} \} \pi_0(\lambda^{(t)}) 
			\textup{N}\bigl [\lambda^* ;   \lambda^{(t)} + \tau M \tilde\nabla \log \Pi\{\lambda^{(t)} \mid y\},  2\tau M \bigr] }.
		\)
		Otherwise, set $\lambda^{(t+1)} \leftarrow \lambda^{(t)}$, $z^{(t+1)} \leftarrow z^{(t)}$.
	\end{itemize}
	In the above, $M\in\mathbb{R}^{d\times d}$ is positive definite and $\tau>0$ is the step size.

\section{Asymptotic Theory}

Many Bayesian models satisfy a Bernstein-von Mises (BvM) theorem under suitable regularity conditions, that is the posterior distribution of $\sqrt{n}(\lambda-\lambda_n)$ where $\lambda_n$ denotes the maximum likelihood estimator (MLE) converges to a normal distribution centered at $0$, with covariance equal to the inverse Fisher information evaluated at $\lambda_0$, denoted by $H^{-1}_0$.

In a canonical Bayesian approach involving latent variable $\zeta$ (that is not conditionally determined), one could focus on the ``integrated posterior'' based on integrated likelihood \citep{berger1999integrated,severini2007integrated}, $\Pi(\lambda\mid y)  \propto \bigl\{ \int L(y, \textup{d} \zeta; \lambda )\bigr\}\pi_0(\lambda)$. For this integrated posterior, BvM results hold for $\sqrt{n}(\lambda-\lambda_n)$ under appropriate conditions, with asymptotic covariance  $H^{-1}_0$ \citep{bickel2012semiparametric,castillo2015bernstein}.

Since $z$ is now conditionally determined given $(y,\lambda)$ under our bridged posterior, it may seem intuitive to expect that the posterior of $\lambda$ would reflect a lower amount of uncertainty (such as having smaller marginal variances) compared to its integrated posterior counterpart. Surprisingly, we dispel this belief  in the asymptotic regime---our result below proves that the bridged posterior of $\lambda$ enjoys the same BvM result with covariance  $H^{-1}_0$.

We establish sufficient conditions for  BvM results under both parametric and semi-parametric cases. To be clear, the parametric setting commonly refers to when both $\lambda$ and $z$ have fixed dimensions, while the semi-parametric one does to when  $\lambda$ has a fixed dimension, but $z$ has a dimension that could grow indefinitely (for instance, increasing with $n$). Therefore, the result developed under the semi-parametric setting can be easily extended to the parametric setting, under the same sufficient conditions while fixing the dimension of $z$.

In the following, we first focus on the BvM result for general bridged posterior which may or may not be based on a profile likelihood. Because we consider a broad family of distributions, we rely on relatively strong conditions here, such as  differentiability of the likelihood in a parametric setting. Next, we relax the differentiability requirements and extend our scope to the semi-parametric setting. As this latter setting presents more challenging conditions, we will restrict our focus to the sub-class of bridged posteriors based on profile likelihoods in our treatment of the semi-parametric case.

In both settings, we consider $\lambda$ in the parameter space $\Theta \subset \mathbb{R}^{d}$ and that there is a fixed ground-truth $\lambda_0$, and the prior density $\pi_0(\lambda)$ to be continuous at $\lambda_0$ with $\pi_0(\lambda_0)>0$.  We use  $\|\cdot\|$  as the Euclidean--Frobenius norm, and $B_{r}(\lambda_0) =\{\lambda\in \mathbb{R}^d : \|\lambda-\lambda_0\| <  r\}$ as a ball of radius $r$. 

\subsection{General Bridged Posterior under Parametric Setting}\label{subsec:para}

For a real-valued function $\alpha(x)$ defined on $\mathbb{R}^d$, we denote first, second and third derivatives by $\alpha'(x)\in \mathbb{R}^d$, $\alpha''(x)\in\mathbb{R}^{d\times d}$ and $\alpha'''(x)\in\mathbb{R}^{d\times d \times d}$, respectively. For a vector-valued function $\alpha(x)=\{\alpha_1(x),\dots,\alpha_m(x) \}$, we again use notations $\alpha'(x)$, $\alpha''(x)$ and $\alpha'''(x)$ to denote the derivatives, to be understood as tensors one order higher. We say a sequence of functions $\alpha_n$ uniformly bounded on $E$ if the set $\{\|\alpha_n(x)\|:x\in E,n\in\mathbb{N}\}$ is bounded. We use $\xrightarrow[n\to\infty]{a.s.[y_{1:n}]} $ for almost sure convergence, and  $ \xrightarrow[n\to\infty]{P_{\lambda_0,\zeta_0}}$ for   convergence in probability.

To ease the notation, we define $$l_n(\lambda, \zeta) = \log L(y_{1:n}, \zeta ;\lambda)/n, \qquad  \hat{l}_n(\lambda) = l_n\{\lambda, \hat{z}_n(\lambda)\} = \log L\{ y_{1:n}, \hat{z}_n(\lambda); \lambda\} / n,$$
where the CDLV $\hat{z}_n(\lambda) :=
\arg\min_{\zeta} g_n(\zeta, y_{1:n};\lambda)$. Let $E$ be an open and bounded subset of $\Theta$ such that $\lambda_0\in E$. We first state and explain some assumptions. 
\renewcommand{\theenumi}{A\arabic{enumi}}
\renewcommand{\labelenumi}{(\theenumi)\;}
\begin{enumerate}[leftmargin=1cm]
	\item \label{assump:1-1}   The function $l_n$ has continuous third derivatives on $E \times \hat{z}_n(E)$, $\hat{z}_n$ has continuous third derivatives on $E$, $l_n'''$ is uniformly bounded on $E \times \hat{z}_n(E)$, and $\hat{z}_n'''$ is uniformly bounded on $E$, $a.s.[y_{1:n}]$.
    \item \label{assump:1-2} The two functions  $\hat{z}_n\to \hat{z}_*$  $a.s.[y_{1:n}]$ on $\Theta$ for some function $\hat{z}_*$, $l_n\to l_*$  $a.s.[y_{1:n}]$ for some function $l_*$.
    \item \label{assump:1-3} The limit $l_*$ has positive definite $-l_*''\{\lambda_0, \hat{z}_*(\lambda_0)\}$ and satisfies $\frac{\partial l_*(\lambda_0, \zeta)}{\partial \zeta}|_{\zeta=\hat{z}_*(\lambda_0)}=0$.
    \item \label{assump:1-4}  For some compact $K \subseteq E$ with $\lambda_{0}$ in the interior of $K$,
    \(
    l_*(\lambda, \zeta) < l_*\{ \lambda_{0},\hat{z}_*(\lambda_0) \} \text{ for all } \lambda \in K \backslash\{\lambda_{0}\}, \zeta \in \hat{z}_*(E) \ a.s.[y_{1:n}],
    \) 
    \(
    \limsup_{n} \sup_{\lambda \in \Theta \backslash K, \zeta \in \hat{z}_n(\Theta)} l_n( \lambda,\zeta) 
    < l_*\{ \lambda_{0}, \hat{z}_*(\lambda_0) \} \ a.s.[y_{1:n}].
    \)
\end{enumerate}
Conditions (\ref{assump:1-1}--\ref{assump:1-2})	are often imposed to enable a second-order Taylor expansion \citep{miller2021asymptotic}; (\ref{assump:1-3}) focuses on the cases when $\lambda=\lambda_0$  and gives the local second-order optimal condition of $l_*(\lambda_0, \zeta)$ at $\zeta=\hat z_*(\lambda_0)$, where $\hat z_*(\lambda_0)$ can be produced as the minimizer of another loss function $g$; (\ref{assump:1-4}) ensures the dominance of $l_*$ at $\{\lambda_0, \hat z_*(\lambda_0) \}$ over all possible $(\lambda, \zeta)$ in the described neighborhood, including those points with $\lambda\neq \lambda_0$. With the above, we are ready to state the BvM result on the general bridged posterior for parametric models where $\zeta \in \mathbb{R}^{p}$ has a fixed and finite dimension.

\begin{theorem}\label{thm:BvM_parametric}
	Under (\ref{assump:1-1}--\ref{assump:1-4}), there is a sequence $\lambda_n\to\lambda_0$ such that $\hat{l}_n'(\lambda_n)=0$ for all $n$ large enough, $\hat{l}_n(\lambda_n)\to \hat{l}_*(\lambda_0)$ where $\hat{l}_*(\lambda) = l_*\{ \lambda,\hat{z}_*(\lambda) \}$. Further,  letting $q_{n}$ be the density of $\sqrt{n}(\lambda-\lambda_{n})$ when $\lambda \sim \Pi_n(\lambda \mid y)$, and $\mathcal N$ the normal density, we have the total variational distance $\text{d}_{\text{TV}}\bigl\{q_n,\mathcal{N}\bigl(0,H_0^{-1}\bigr)\bigr\} \xrightarrow[n\to\infty]{a.s.[y_{1:n}]} 0$ with $H_0=\hat{l}_*''(\lambda_0)$.
\end{theorem}

The result above shows that fixing $\zeta$ to $z$ does not impact the asymptotic variance of $\lambda$. On the other hand, since $z$ is finite-dimensional and differentiable on $E$, we can use the delta method to find out the asymptotic covariance of $z$. For bridged posterior using profile likelihood, we do find lower uncertainty in $\Pi(z\mid y)$ under a bridged posterior compared to $\Pi(\zeta\mid y)$ under an integrated one, as formalized below.

\begin{corollary}\label{coro:smallvar}
	Under (\ref{assump:1-1}--\ref{assump:1-4}) and $g_n(\zeta, y_{1:n};\lambda) = -L(y_{1:n},\zeta; \lambda)$, for $j=1,\ldots,p$, the asymptotic variance of the $j$-th element of $\sqrt{n}\{\zeta- \hat{z}_n(\lambda_{n})\}$ is strictly greater than the one of the $j$-th element of $\sqrt{n}\{\hat{z}_n(\lambda)-\hat{z}_n(\lambda_{n})\}$.
\end{corollary}

\begin{remark}
    {In the proof of Corollary \ref{coro:smallvar}, we show that for the bridged posterior based on profile likelihood, the inverse asymptotic variance $H_0= -\hat{l}''_*(\lambda_0)= -l_{*,\lambda_0 \lambda_0} + l_{*,\lambda_0 \zeta_0}l_{*,\zeta_0\zeta_0}^{-1} l_{*,\zeta_0\lambda_0}$, 
    where $l_{*,\lambda_0\lambda_0}, l_{*,\zeta_0\zeta_0}, l_{*,\zeta_0\lambda_0}, l_{*,\lambda_0 \zeta_0}$ are the second partial derivatives of $l_*$ evaluated at $\lambda=\lambda_0,\zeta=\zeta_0=\hat{z}_*(\lambda_0)$. On the other hand, for the integrated posterior (marginal posterior), letting $\tilde Q_n$ denote the posterior distribution of  $\sqrt{n} (\lambda -\lambda_n)$ when $\lambda \sim \tilde \Pi(\lambda \mid y) \propto \pi_0(\lambda) \int L(y, \textup{d} \zeta; \lambda)$, the BvM theorem \citep{miller2021asymptotic} states that $\text{d}_{\text{TV}}\bigl\{\tilde Q_n,\mathcal{N}\bigl(0,\tilde H_0^{-1}\bigr)\bigr\} \xrightarrow[n\to\infty]{a.s.[y_{1:n}]} 0$ where $\tilde H_0^{-1}$ is the $\lambda$-block of the inverse full Fisher information $-\bigl[{l}_*''(\lambda_0,\zeta_0)\bigr]^{-1}$. Block matrix inversion then shows that $H_0 = \tilde H_0$.}
\end{remark}

\subsection{Bridged Posterior using Profile Likelihood under Semi-parametric Setting}

In the semi-parametric setting, we assume that $\zeta$ can be infinite-dimensional and live in some Hilbert space $\mathcal{H}$, and that there exists a fixed $\zeta_0\in\mathcal{H}$. We define
$$
l_n(\lambda, \zeta) = \log L(y_{1:n}, \zeta ;\lambda)/n,\qquad \hat{l}_n(\lambda)
= \log \{\sup_ \zeta  L(y_{1:n}, \zeta ;\lambda)\} / n,
$$
where the former corresponds  to  a full likelihood $L(y_{1:n}, \zeta ;\lambda)$ with unconstrained $\zeta$, and the latter to a profile likelihood $\sup_\zeta  L(y_{1:n}, \zeta ;\lambda)$. 
In addition to the potentially infinite dimension, another challenge is that $l_n(\lambda,\zeta)$ may not be differentiable with respect to $\zeta$. 

To facilitate analysis under these challenges, we use the ``approximately least-favorable submodel'' technique;   \cite{kosorok2008introduction} provides a detailed explanation. For this section to be self-contained, we overview the important definitions that are involved as the building blocks for establishing BvM results.

\noindent\textbf{Submodel:} For each  $(\lambda, \zeta)\in \Theta \times \mathcal H$, consider a map  $\tilde \zeta_t(\lambda,\zeta)$ indexed by $t\in \Theta\subset \mathbb{R}^d$, such that
\[\label{eq:z_map}
l_n \{ t, \tilde \zeta_{t}(\lambda, \zeta)\} \text{ is twice differentiable in }t\in \Theta,  \qquad \tilde \zeta_{t=\lambda}(\lambda, \zeta) = \zeta.
\]
Commonly, $l_n\{ t, \tilde \zeta_{t}(\lambda, \zeta)\}$ is called a ``submodel'' with parameters $(t,\lambda,\zeta)$ \citep{murphy2000profile}. For convenience, we use notation $\tilde{l}_n(t, \lambda, \zeta):=l_n\{ t, \tilde \zeta_{t}(\lambda, \zeta)\}$.

\noindent\textbf{Efficient score and Fisher information:} Conventionally, the $\lambda$-score function of the full likelihood is $\dot{l}_n (\lambda, \zeta)=\frac{\partial l_n(\lambda,\zeta)}{\partial \lambda}$.
Consider a direction $\delta \in \widetilde{\mathcal{H}}$ (another Hilbert space) such that a path $\{\zeta_\gamma^\delta \in \mathcal H\}_{\gamma\in \mathbb{R}^d}$ with $\zeta_\gamma^\delta\to \zeta_0$ as $\gamma\to \lambda_0$.
We can now define the generalized $\zeta$-score function at $\zeta=\zeta_0$ in the direction of $\delta$ by $A^n_{\lambda_{0}, \zeta_{0}} \delta := \dfrac{\partial l_n(\lambda_{0}, \zeta_{\gamma}^\delta)}{\partial \gamma}\bigg|_{\gamma=\lambda_0}$, where $A^n_{\lambda_{0}, \zeta_{0}}:\widetilde{\mathcal{H}} \mapsto L_2^d(P_{\lambda_0,\zeta_0})$ is a map, and $L_2^d(P_{\lambda_0,\zeta_0})$ is the space of $d$-dimensional vector-valued functions $\{\alpha_1(y),\dots,\alpha_d(y)\}$ where each $\alpha_i(y)$ is $L_2$-integrable on $y\sim P_{\lambda_0,\zeta_0}$. In the appendix, we provide an illustration of the above via the Cox regression model.

The ``efficient score function" for $\lambda$ at $(\lambda_0,\zeta_0)$ is defined by
\(
&\mathscr{Q}\dot{l}_n(\lambda_0,\zeta_0) := \dot{l}_n (\lambda_0, \zeta_0) - \mathscr{P}\dot{l}_n (\lambda_0, \zeta_0),\\
&\mathscr{P}\dot{l}_n (\lambda_0, \zeta_0) := {\arg\min}_k \mathbb{E}_{\lambda_0,\zeta_0} \big\|\dot{l}_n (\lambda_0, \zeta_0) - \kappa\big\|^2, \;\; \kappa\in \text{closed linear span of } A^n_{\lambda_{0}, \zeta_{0}}\delta.
\)
The ``efficient Fisher information" at $(\lambda_0,\zeta_0)$ is defined as
\(
\tilde{I}_0 := \mathbb{E}_{\lambda_0,\zeta_0} \bigl\{  \mathscr{Q}\dot{l}_n (\lambda_0, \zeta_0)\mathscr{Q}\dot{l}_n (\lambda_0, \zeta_0)^\top\bigr\}.
\)
Equivalently, $\mathscr{P}\dot{l}_n (\lambda_0, \zeta_0)$ is the projection of the score function for $\lambda_0$ onto the closed linear space spanned by the set $\{A^n_{\lambda_{0}, \zeta_{0}} \delta\}_{\delta\in \widetilde{\mathcal{H}}}$.

\noindent\textbf{Least favorable model:} To connect the two topics above, notice that if \eqref{eq:z_map} further satisfies
\(
\frac{\partial \tilde{l}_n(t, \lambda_0, \zeta_0)}{\partial t}\bigg|_{t=\lambda_0} = \mathscr{Q}\dot{l}_n(\lambda_0, \zeta_0),
\)
then we have the submodel $\tilde{l}_n(t, \lambda, \zeta)$  ``least favorable" at  $t= \lambda_0$.  This is because among all submodels $\tilde{l}_n(t, \lambda_0, \zeta_0)$, this submodel has the smallest Fisher information on each dimension of $t\in\Theta\subset\mathbb{R}^d$ by the definition of $\mathscr{Q}\dot{l}_n(\lambda_0, \zeta_0)$. {Since our focus is on the asymptotic regime, we only need the least favorable model condition to hold in a limiting sense. This leads to the ``approximately least favorable model''.} 
With these ingredients, we are ready to derive our results. We first show that the profile $\hat{l}_n(\lambda)$ is locally asymptotically normal (LAN). We require the following sufficient conditions.

\renewcommand{\theenumi}{B\arabic{enumi}}
\renewcommand{\labelenumi}{(\theenumi)\;}
\begin{enumerate}[leftmargin=1cm]  \setcounter{enumi}{0}
	\item \label{assump:2-1}    There exists a neighborhood $V\subset \Theta \times \Theta \times \mathcal H$ containing $(\lambda_0,\lambda_0,\zeta_0)$ such that
	\begin{itemize}
		\item  $\sup_{(t,\lambda,\zeta)\in V} \big\|\frac{\partial^2 \tilde{l}_n(t, \lambda, \zeta)}{\partial t^2} + H_0\big\| \xrightarrow[n\to\infty]{P_{\lambda_0,\zeta_0}} 0$ for some symmetric $H_0\in \mathbb{R}^{d\times d}$; and 
		\item $\sup_{(t,\lambda,\zeta)\in V}  \sqrt{n}\big\| \frac{\partial \tilde{l}_n(t, \lambda, \zeta)}{\partial t} - \mathbb{E}_{\lambda_0,\zeta_0} \frac{\partial \tilde{l}_n(t, \lambda, \zeta)}{\partial t}  - h_n\| \xrightarrow[n\to\infty]{P_{\lambda_0,\zeta_0}} 0$ for a sequence of random variables $h_n$ $\in \mathbb{R}^d$,
	\end{itemize}
	where the $P_{\lambda_0,\zeta_0}$ and $\mathbb{E}_{\lambda_0,\zeta_0}$ are defined with respect to the ground-truth distribution of $y_{1:n}$.
	\item The function $\hat{z}_n(\lambda)$ converges to $\zeta_0$ when $\lambda\to\lambda_0$ and $n\to\infty$.
	\item \label{assump:2-3} There exists a neighborhood $U\subset \Theta$ containing $\lambda_0$ such that
	\[
	\mathbb{E}_{\lambda_0,\zeta_0} \frac{\partial \tilde{l}_n\{t, \lambda, \hat{z}_n(\lambda)\}}{\partial t}\bigg|_{t=\lambda_0} = o_{P_{\lambda_0,\zeta_0}}(1)\bigl(\|\lambda-\lambda_0\| + n^{-1/2} \bigr)
	\label{eq:expect_score}
	\]
	holds for all $\lambda\in U$. Here, $o_{P_{\lambda_0,\zeta_0}}(1)$ refers to a term that converges to $0$ in $P_{\lambda_0,\zeta_0}$ as $n\to \infty$.
\end{enumerate}
Conditions (\ref{assump:2-1}--\ref{assump:2-3}) are the approximately least favorable submodel conditions.
A similar result for iid data given $(\zeta_0,\lambda_0)$
has been previously shown in \citet[Theorem 1]{murphy2000profile}. However, our result is more general and holds regardless of whether { $(y_{1:n}\mid \zeta_0,\lambda_0)$  are iid or not, and may be of independent interest outside the context of establishing BvM results}.

\begin{lemma}\label{lemma:LAN}
    Under (\ref{assump:2-1}--\ref{assump:2-3}), there exists a neighborhood $B_{\epsilon}(\lambda_0)$ for some $\epsilon>0$ such that
    \bel
	\hat{l}_{n}(\lambda)- \hat{l}_{n}(\lambda_0) ={}  (\lambda - \lambda_0)^\top h_n -\frac{1}{2}  (\lambda - \lambda_0)^\top H_0(\lambda-\lambda_0) 
	 + o_{P_{\lambda_0,\zeta_0}}(1)\Bigl\{\bigl(\|\lambda-\lambda_0\| + n^{-1/2} \bigr)^2\Bigr\} \label{eq:lan}
    \eel
    holds for all $\lambda\in B_{\epsilon}(\lambda_0)$.
\end{lemma}

With the LAN condition for $\hat{l}_n(\lambda)$, we make the probability statement as the BvM result.
\begin{theorem} \label{thm:BvM_semi}
    Assume \eqref{eq:lan} holds with positive definite $H_0$. Suppose that the maximum likelihood estimator $\hat \lambda_n$ exists and converges to $\lambda_0$ when $n\to\infty$; for any $\epsilon >0$, there exists $\delta>0$ such that
    \be
	P_{\lambda_0,\zeta_0} \biggl[ \inf_{  \|\lambda - \hat\lambda_n \| \geq \epsilon} \bigl\{ \hat{l}_n( \hat \lambda_n)-\hat{l}_n(\lambda) \bigr\} \ge  \delta  \biggr] \xrightarrow[n\to\infty]{} 1.
    \ee
    Then letting $\pi_n$ be the density of $\lambda$ when $\lambda \sim \Pi_n(\lambda \mid y)$, we have
    \(
    \int_{B_{\epsilon}(\lambda_{0})} \pi_{n}(\lambda)\, \textup{d} \lambda \xrightarrow[n\to\infty]{P_{\lambda_0,\zeta_0}}  1 \text { for all } \epsilon>0,
    \)
    and letting $q_{n}$ be the density of $\sqrt{n}(\lambda-\hat\lambda_{n})$, we have $\text{d}_{\text{TV}}\bigl\{q_n,\mathcal{N}\bigl(0,H_0^{-1}\bigr)\bigr\} \xrightarrow[n\to\infty]{P_{\lambda_0,\zeta_0}} 0$.
\end{theorem}

\begin{remark}
    We provide a detailed comparison with existing BvM results on semi-parametric models in \cref{sec:compare_bvm}. 
    {
    Here, we compare the asymptotic variance of the bridged posterior with that of the integrated posterior. According to the BvM theorem for the integrated posterior \citep[Theorem 2.1]{bickel2012semiparametric}, under certain regularity conditions, letting $\tilde Q_n$ denote the posterior probability distribution of $\sqrt{n}(\lambda - \lambda_0)$ where $\lambda \sim \tilde \Pi(\lambda \mid y) \propto \pi_0(\lambda) \int L(y, \textup{d} \zeta; \lambda) $, we have $\text{d}_{\text{TV}}\bigl\{\tilde Q_n, \mathcal{N}\bigl(\tilde \Delta_n,\tilde{I}_{0}^{-1}\bigr)\bigr\} \xrightarrow[n\to\infty]{P_{\lambda_0,\zeta_0}} 0$, where $\tilde \Delta_n = \sqrt{n} \tilde{I}_0^{-1} \mathscr{Q}\dot{l}_n(\lambda_0,\zeta_0)$. Here, $\tilde{I}_0$ and $\mathscr{Q}\dot{l}_n(\lambda_0,\zeta_0)$ denote the efficient Fisher information and the efficient score function, respectively, as defined earlier. A key assumption in their theory is the existence of a least favorable model, which ensures that the submodel satisfies 
    $
    \partial \tilde{l}_n(t, \lambda_0, \zeta_0) / {\partial t}|_{t=\lambda_0} = \mathscr{Q}\dot{l}_n(\lambda_0, \zeta_0),
    $
    and under this condition, the efficient Fisher information is given by $\tilde{I}_0 = -\mathbb{E}_{\lambda_0,\zeta_0} \bigl\{ \partial^2 \tilde{l}_n(t, \lambda_0, \zeta_0) / {\partial t^2}|_{t=\lambda_0} \bigr\}$. This formulation coincides with our assumption in (\ref{assump:2-1}), under which we also have inverse asymptotic variance $H_0=\tilde{I}_0$. 
    }
\end{remark}

\section{Numerical Experiments \label{sec:numeric}}

\subsection{Latent Quadratic Exponential Model}\label{subsec:simulation_latent}

We begin our empirical study via simulated experiments comparing a latent normal model and a latent quadratic exponential model, based on Example \ref{example:latent}. To simulate data for benchmarking, we generate random locations $x_1,\dots,x_{1000}\sim \textup{Uniform}(-6,6)$, and ground-truth mean from a latent curve $\tilde z_i=\cos(x_i)$. At each $x_i$, we generate a binary observation
$y_i\sim \textup{Bernoulli}(1 / \{1+\exp(-\tilde z_i)\})$.

We fit the latent quadratic exponential model \eqref{eq:latent_quadratic} and the latent normal model \eqref{eq:latent_normal} to the simulated data. For both models, we assign half-normal $\textup{N}_+(0,1)$ prior on $\tau$ and Inverse-Gamma$(2,5)$ prior on $b$. We use random walk Metropolis for the latent quadratic exponential model, and data augmentation Gibbs sampler for the latent normal model (detail provided in \cref{sec:data_aug_latent}).

\begin{figure}[H]
	\begin{minipage}[b]{1\textwidth}
		\begin{subfigure}[t]{0.23\textwidth}
			\centering
			\includegraphics[width=1\linewidth]{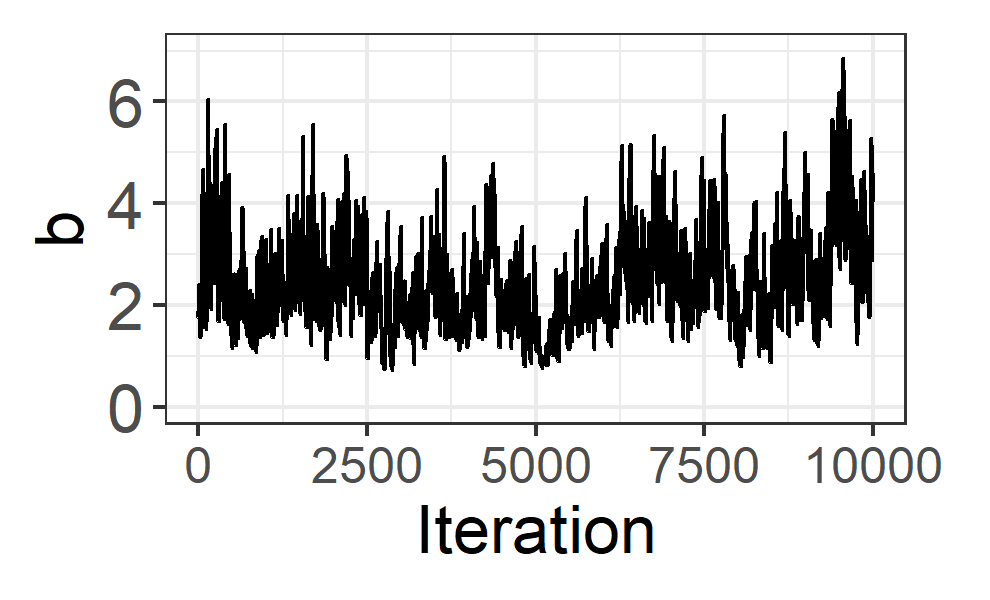}
			\caption*{Traceplot of $b$}
		\end{subfigure}
		\begin{subfigure}[t]{0.23\textwidth}
			\centering
			\includegraphics[width=1\linewidth]{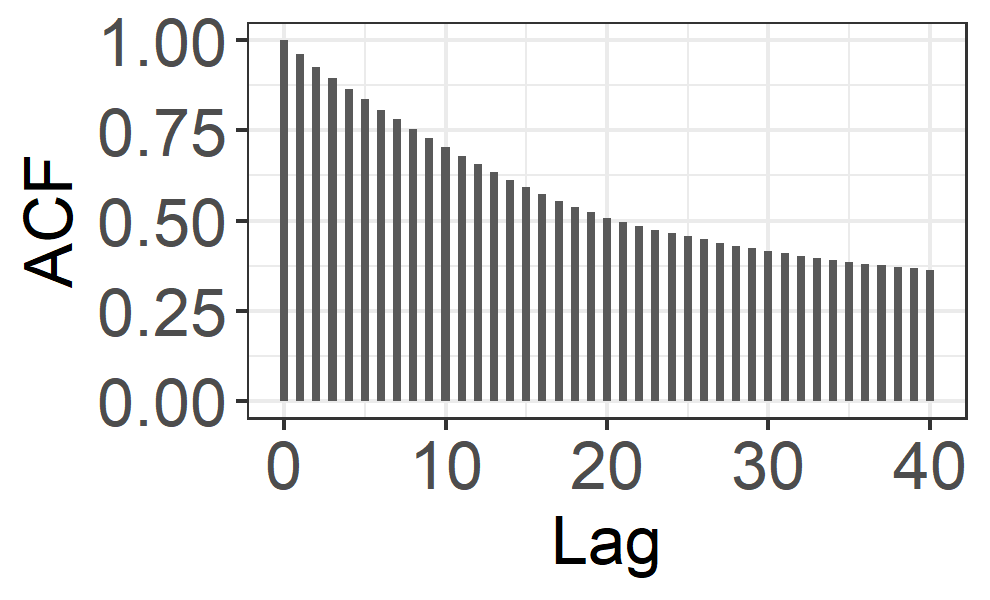}
			\caption*{Autocorrelation of $b$}
		\end{subfigure}
		\begin{subfigure}[t]{0.23\textwidth}
			\centering
			\includegraphics[width=1\linewidth]{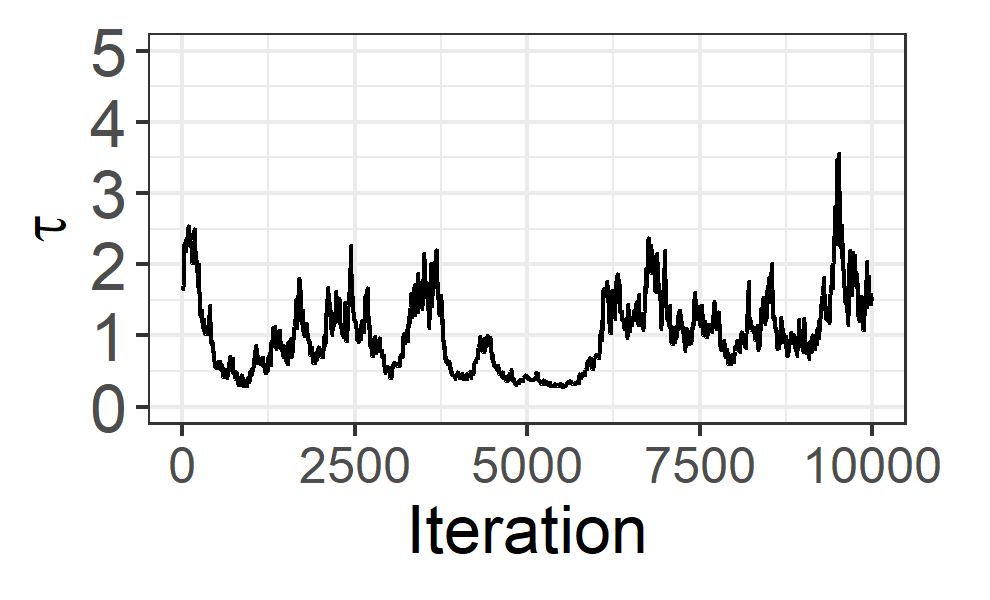}
			\caption*{Traceplot of $\tau$} 
		\end{subfigure}
		\begin{subfigure}[t]{0.23\textwidth}
			\centering
			\includegraphics[width=1\linewidth]{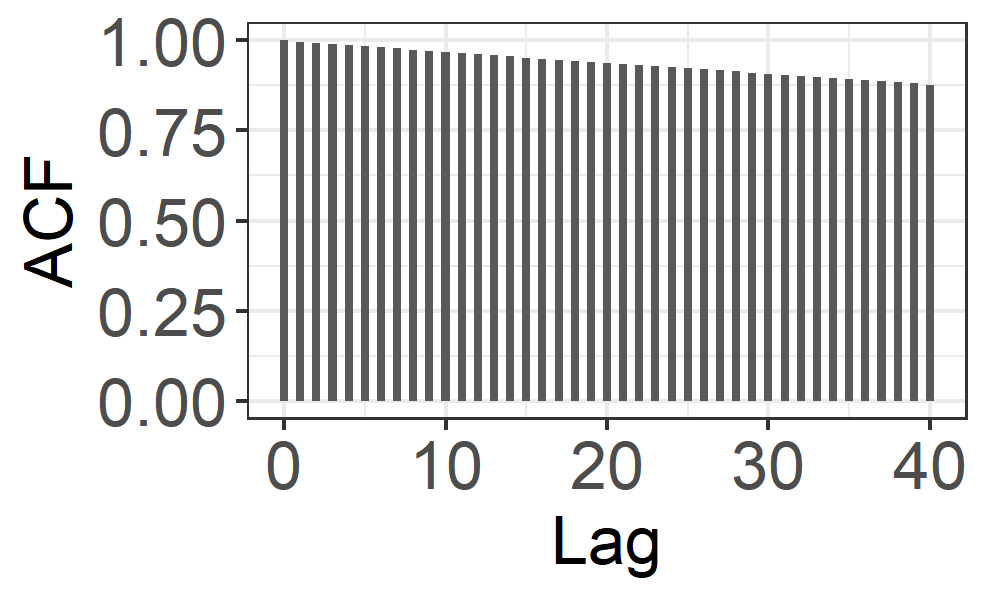}
			\caption*{Autocorrelation of $\tau$} 
		\end{subfigure}\\
		\centering (a) The MCMC samples produced by the data augmentation Gibbs sampler applied on the latent normal model.
	\end{minipage}
	\begin{minipage}[b]{1\textwidth}
		\begin{subfigure}[t]{0.23\textwidth}
			\centering
			\includegraphics[width=1\linewidth]{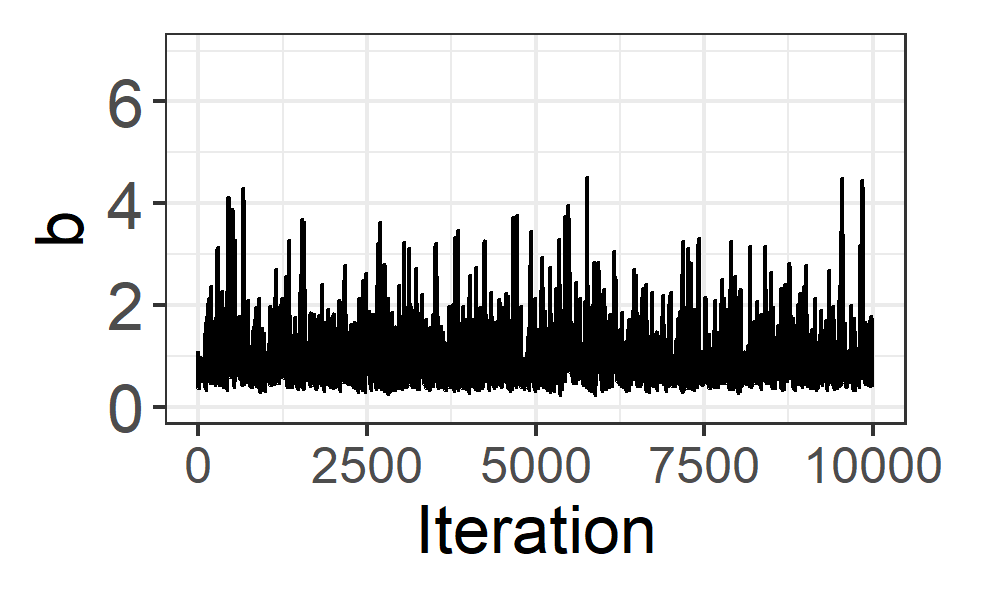}
			\caption*{Traceplot of $b$}
		\end{subfigure}
		\begin{subfigure}[t]{0.23\textwidth}
			\centering
			\includegraphics[width=1\linewidth]{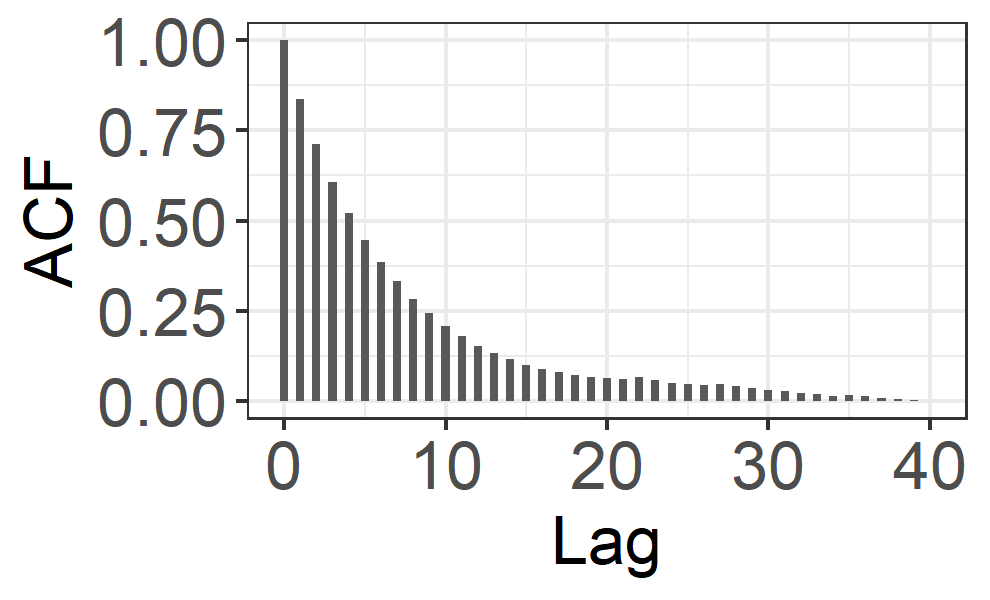}
			\caption*{Autocorrelation of $b$}
		\end{subfigure}
		\begin{subfigure}[t]{0.23\textwidth}
			\centering
			\includegraphics[width=1\linewidth]{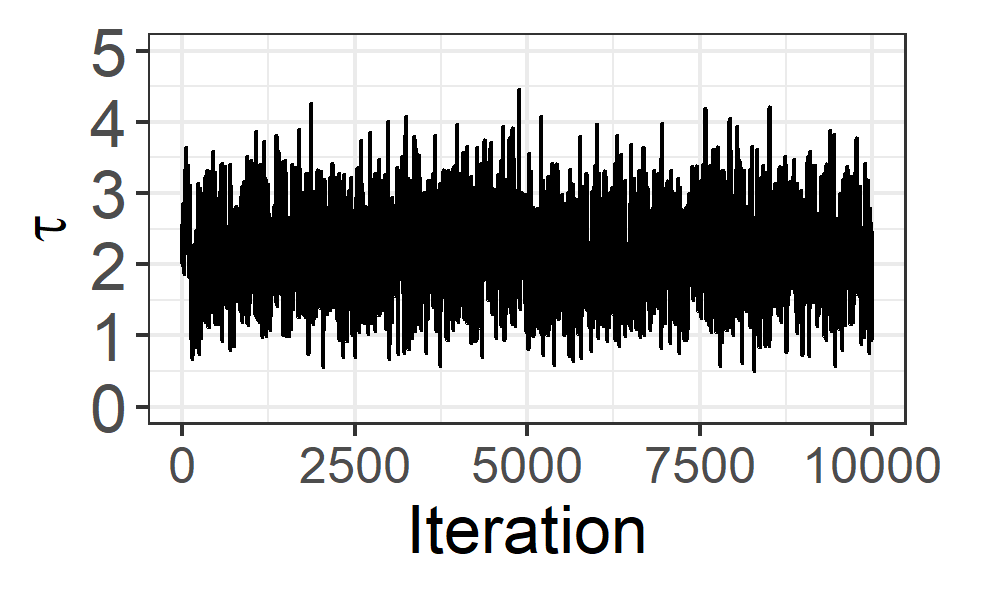}
			\caption*{Traceplot of $\tau$} 
		\end{subfigure}      
		\begin{subfigure}[t]{0.23\textwidth}
			\centering
			\includegraphics[width=1\linewidth]{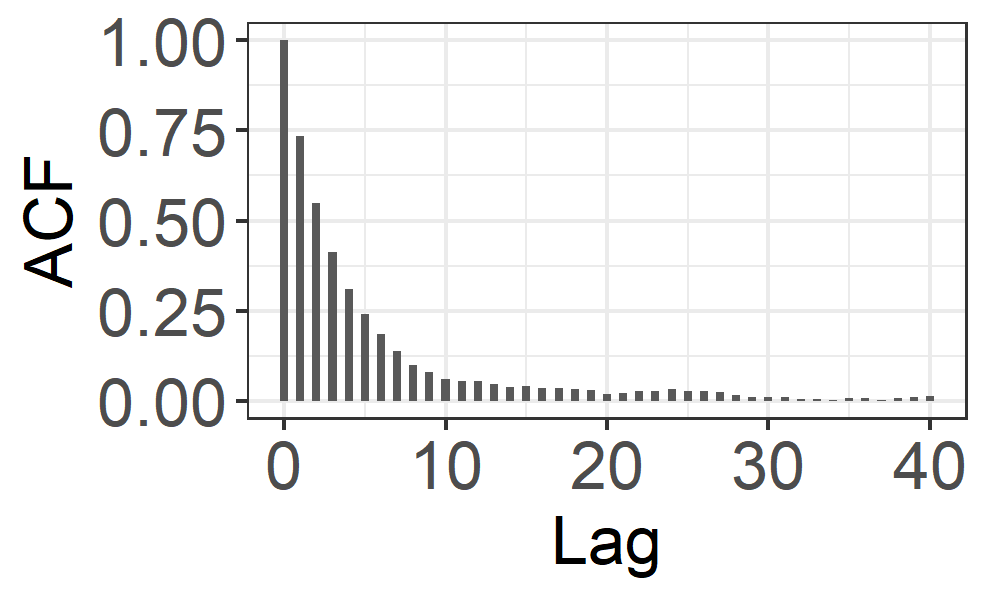}
			\caption*{Autocorrelation of $\tau$} 
		\end{subfigure}\\
		\centering (b) The MCMC samples produced by the random-walk Metropolis applied on the latent quadratic exponential model.
	\end{minipage}
	\caption{Compared to the latent normal distribution using data augmentation Gibbs sampler, the latent quadratic exponential model (a bridged posterior model) can be estimated using a much simpler random-walk Metropolis, while enjoying faster mixing of the Markov chains.
		\label{fig:computing_perf}}
\end{figure}

We run each MCMC algorithm for $10,000$ iterations and discard the first $2,000$ as burn-ins. The latent quadratic exponential model takes about $8.4$ minutes, and the latent normal model takes about $11.9$ minutes on a $12$-core processor. Figure \ref{fig:computing_perf} compares the mixing of MCMC algorithms for those two models. Clearly, the latent quadratic exponential model mixes better, while taking less runtime. In terms of effective sample size for $(b,\tau)$ per time unit (10 seconds wall time) (ESS/time), the latent quadratic exponential model achieves $0.059$ for $b$ and $0.106$ for $tau$, while the latent normal model yields only $0.0087$ and $0.0008$, respectively.

\begin{figure}[H]
	\centering
	\begin{subfigure}[t]{0.45\textwidth}
		\centering
		\includegraphics[width=1\linewidth]{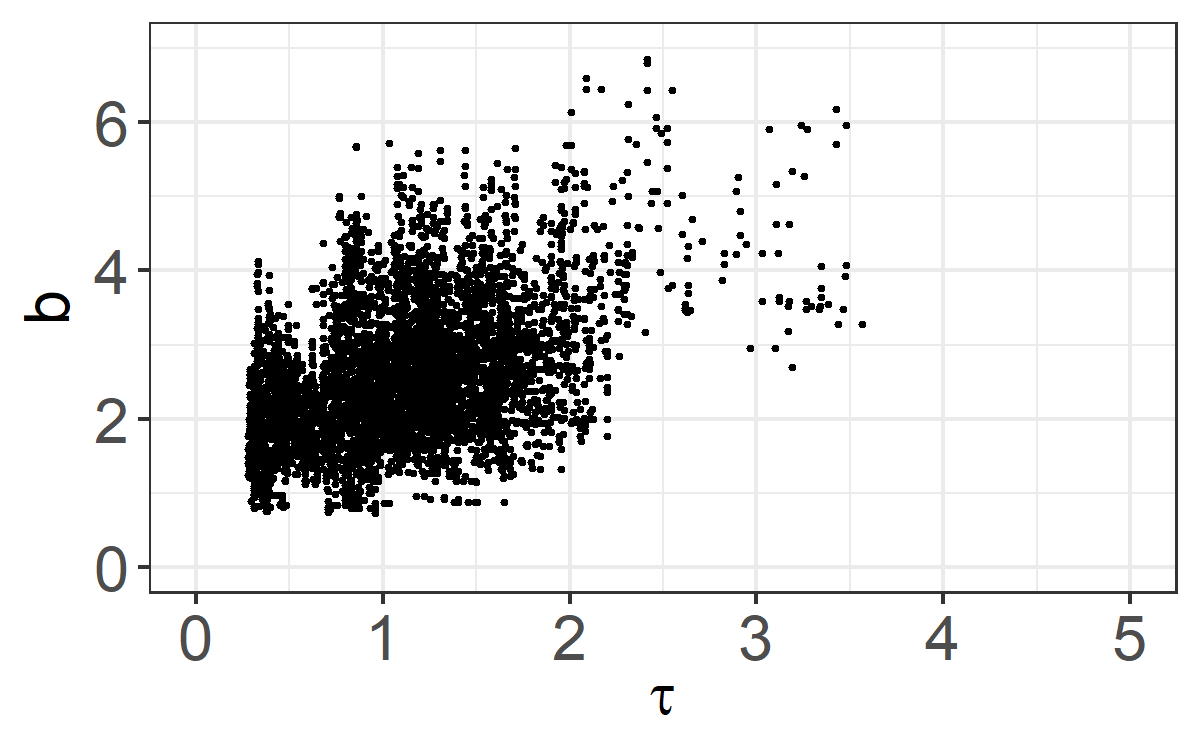}
		\caption{Posterior distribution of $(b,\tau)$  from latent normal model.} 
	\end{subfigure}
	\begin{subfigure}[t]{0.45\textwidth}
		\centering
		\includegraphics[width=1\linewidth]{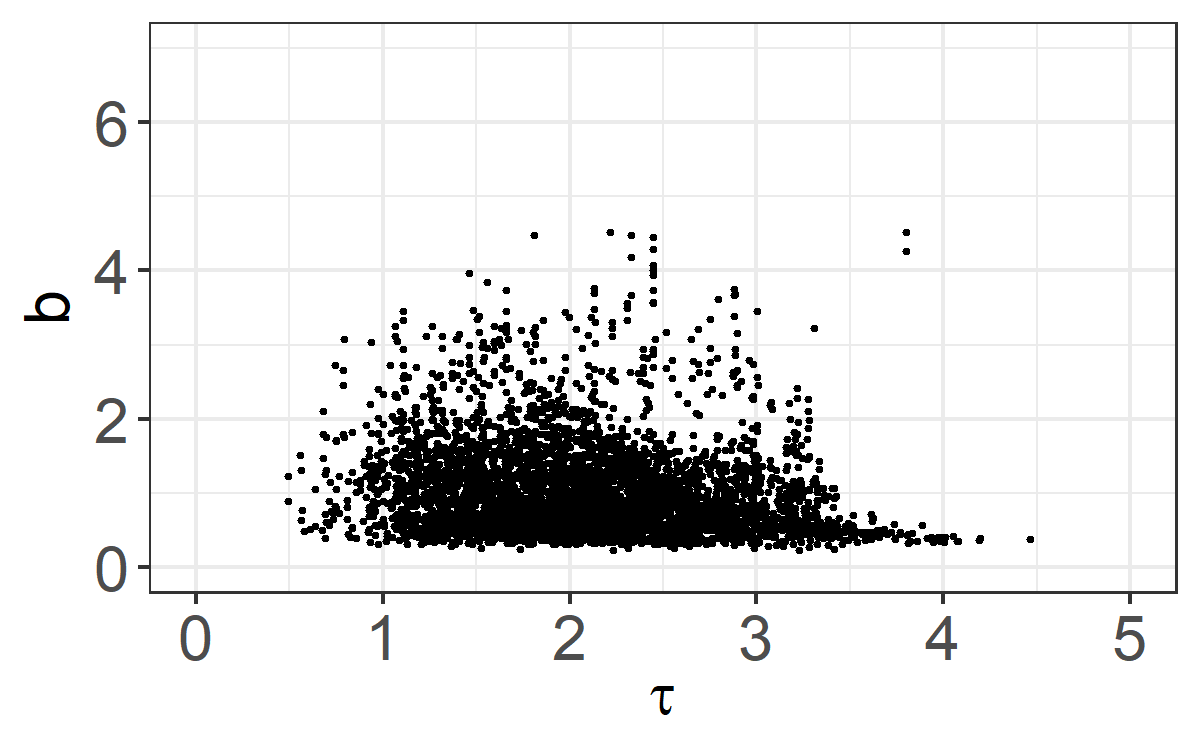}
		\caption{Posterior distribution of $(b,\tau)$ from latent quadratic exponential model.}
	\end{subfigure}
	\centering
	\begin{subfigure}[t]{0.45\textwidth}
		\centering
		\includegraphics[width=\linewidth]{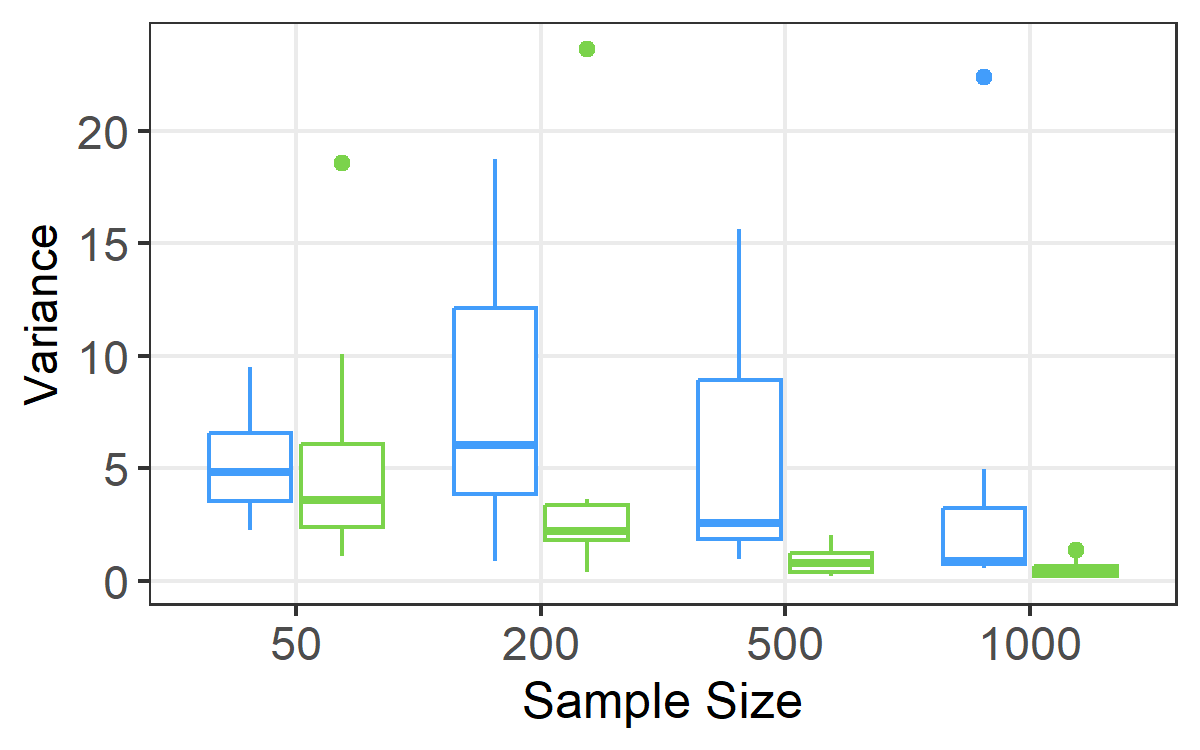}
		\caption{Boxplots of posterior variances of $b$ at different sample sizes.}
	\end{subfigure}
	\begin{subfigure}[t]{0.45\textwidth}
		\centering
		\includegraphics[width=\linewidth]{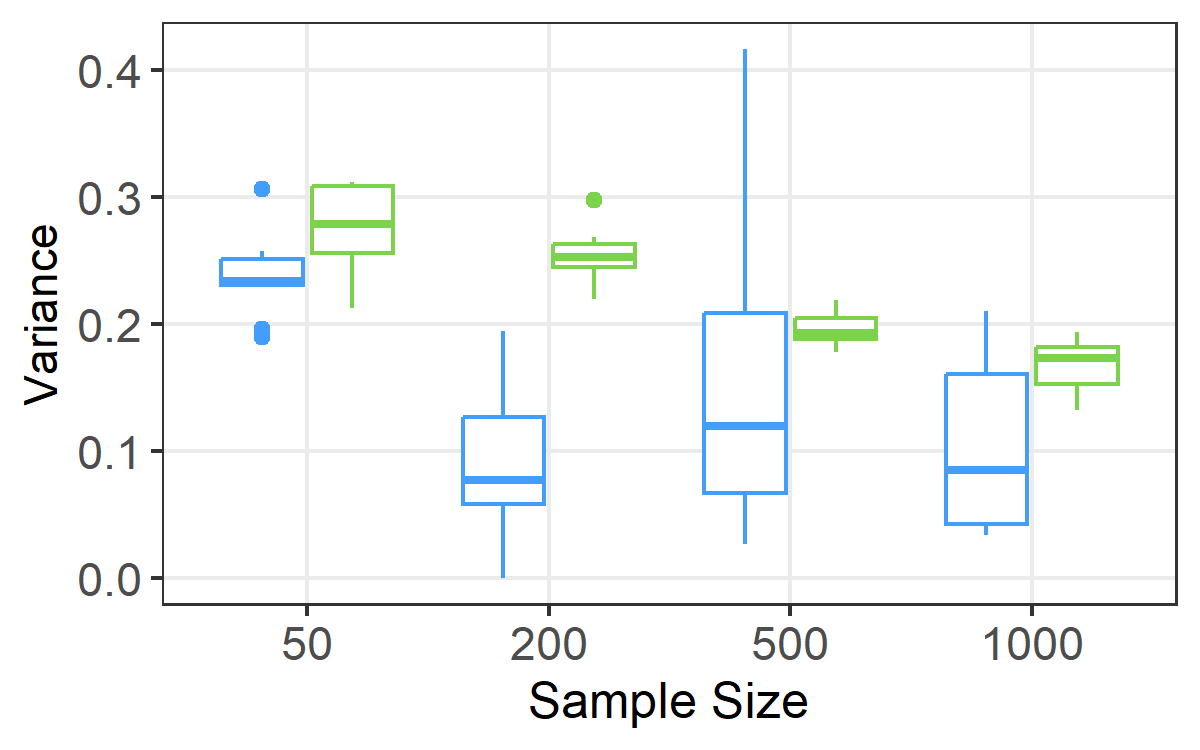}
		\caption{Boxplots of posterior variances of $\tau$  at different sample sizes.}
	\end{subfigure}
	\caption{The posterior distributions of the covariance kernel parameters from the latent normal model (Panel a) and the latent quadratic exponential model (Panel b), collected from two experiments under sample size 1000. The experiments are repeated under different sample sizes, and the posterior variances of $b$ and $\tau$ from the latent quadratic exponential model (green) and the latent normal model (blue) are shown in Panels c and d.
		\label{fig:latent_gau_posterior}}
\end{figure}

Next, we compare the posterior distributions of parameters $(\tau,b)$. As can be seen in Figure \ref{fig:latent_gau_posterior}, these two distributions show  similar ranges of $\tau$ and $b$ in the high posterior probability region. Since these two distributions correspond to two distinct models, we do not expect the distributions of $\tau$ or $b$ to match exactly. On the other hand, we can see that the posterior variances are on the same scale, with $\text{Var}(b\mid y)=0.60^2$ and $\text{Var}(\tau\mid y)=0.65^2$  for the latent normal model, and $\text{Var}(b\mid y)=0.94^2$ and $\text{Var}(\tau\mid y)=0.51^2$ for the latent quadratic exponential model. We repeat the experiments and compare the variances under different sample sizes. Additionally, we compare these empirically to two posterior approximation algorithms, integrated nested Laplace approximation and mean-field variational inference,  with details in \cref{sec:approx_latent} and \cref{sec:sim_latent_normal_var}.

\subsection{Bayesian Maximum Margin Classifier} \label{subsec:missing}
To illustrate the strengths of our approach in terms of uncertainty quantification and borrowing information from unlabeled data, we apply the Bayesian maximum margin classifier (Example \ref{example:svm}) to prediction on heart failure-related deaths. The dataset we consider comprises 299 total patients who had a previous occurrence of heart failure. For each patient, there are 12 measured clinical features, with binary outcomes $y_i$ on whether the patient died during a follow-up care period between April and December 2015, at the Faisalabad Institute of Cardiology and at the Allied Hospital in Faisalabad, Pakistan. There are 194 men and 105 women between age 40 and 95.

\begin{figure}[H]
	\centering
	\includegraphics[width=0.6\linewidth]{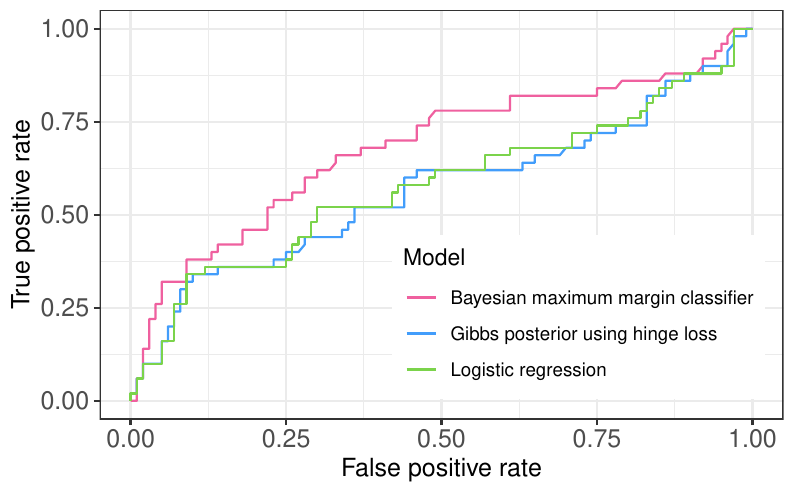}
	\caption{The prediction receiver operating characteristic curves from the three models.
		\label{fig:roc}}
\end{figure}
We mask the outcomes of randomly chosen 97 men and 52 women (corresponding to roughly 50\% missing labels), and fit the data under the (i)  Bayesian maximum margin classifier model, (ii) a Gibbs posterior model using hinge loss, and (iii) logistic regression. We specify the priors {$\lambda \sim \text{Gamma}(3, 2)$} for models (i) and (ii), and $\zeta_w \sim \text{N}(0,3^2 I)$ and $\zeta_b \sim \text{N}(0,3^2)$ for models (ii) and (iii). For each model, we run MCMC for $1,500$ iterations and discard the first $500$ as burn-in. At each iteration, we make a binary prediction on each unlabeled $x_j$, using the average as a posterior estimate for predicting $P(y_j=1\mid y_{1:n})$ for $j=n+1,\ldots, n+k$. Comparing each of these prediction probabilities with the true $y_j$ produces the prediction receiver operating characteristic curves, displayed in Figure \ref{fig:roc}. For binary estimates, we threshold the probability at 0.5 and report classification accuracy. Figure \ref{fig:roc} reveals a barely noticeable difference between logistic regression and the Gibbs posterior using hinge loss. In contrast, the Bayesian maximum margin classifier clearly produces higher area under the curve (AUC). This advantage is also apparent in terms of classification accuracy,  displayed in Table \ref{tab:accuracy_heart_failure}.

To see that these gains are largely due to borrowing information from the unlabeled data, we also fit a support vector machine only using the labeled part, and hold out the unlabeled portion for prediction. Here, the classification accuracy falls to similar levels as the other two Bayesian models, with thresholding probability at 0.5. 

\begin{table}[H]
	\centering
	\caption{Prediction accuracy for heart failure dataset using four methods.}
	\begin{tabular}{lcc}
		\toprule
		Method & Area Under ROC Curve & Classification Accuracy  \\
		\midrule
		Bayesian maximum margin classifier & 0.681 & 0.707 \\
		Gibbs posterior using hinge loss   &  0.568 & 0.653\\
		Support vector machine         & -  & 0.653\\
		Logistic regression   & 0.577 & 0.673 \\
		\bottomrule
	\end{tabular}
	\label{tab:accuracy_heart_failure}
\end{table}

\begin{figure}[H]
	\centering
	\begin{subfigure}{0.45\textwidth}
		\centering
		\includegraphics[width=1\linewidth]{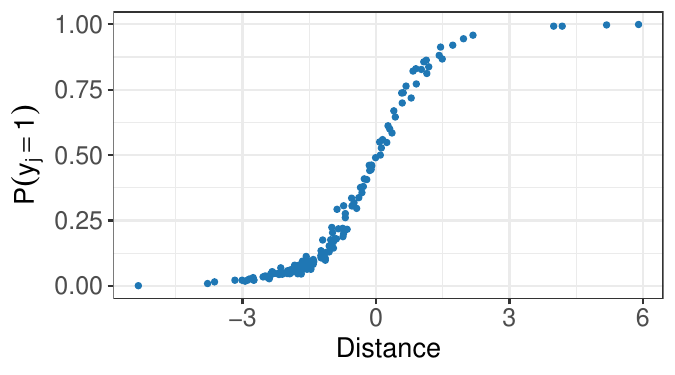}
		\caption{Posterior prediction $P(y_j=1)$ versus distance to the posterior mean of  decision boundary hyperplane. }
	\end{subfigure}
	\begin{subfigure}{0.45\textwidth}
		\centering
		\includegraphics[width=1\linewidth]{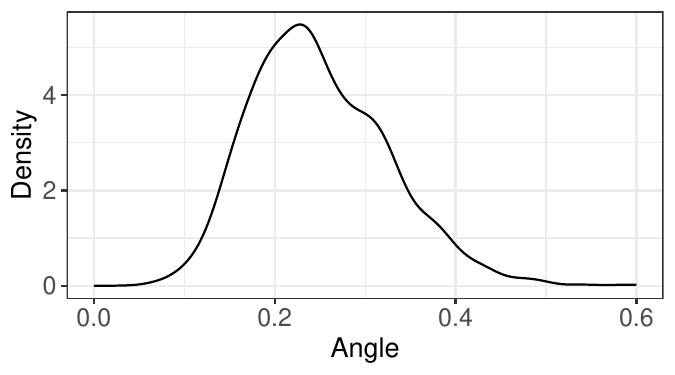}
		\caption{Posterior distribution of the absolute angle (in radians) between $z_w$ and the posterior mean $\bar z_w$.}
	\end{subfigure}
	\caption{Uncertainty estimates for the Bayesian maximum margin classifier applied on the heart failure dataset.} 
	\label{fig:bmmc_uq}
\end{figure}

Finally, in addition to ROC curves, Figure \ref{fig:bmmc_uq} shows the other uncertainty estimates that describe how the posterior prediction $P(y_j=1)$ changes with the distance between $x_j$ and the posterior mean of the decision boundary hyperplane $\{x:z_w ^{\top} x + z_b=0\}$. We can also consider how the posterior distribution describing this decision boundary varies around the posterior mean, in terms of angle between $z_w$ and $\bar z_w$.

\section{Data Application: Harmonization of Functional Connectivity Graphs}

We now use the proposed method to model a collection of raw functional connectivity graphs. The graphs were extracted from resting-state functional magnetic resonance imaging (rs-fMRI) scans, collected from $S=166$ subjects, of whom 64 are healthy subjects and 102 are at various stages of Alzheimer's disease. For each subject, a functional connectivity matrix was produced via a standard neuroscience pre-processing pipeline \citep{ding2006granger}, summarized in the form of a symmetric, weighted adjacency matrix, denoted by $A^{(s)}\in \mathbb{R}_{\ge 0}^{R\times R}$ between $R=116$ regions of interests (ROIs) for subjects $s=1,\ldots, S$; there are no self-loops---that is, $A^{(s)}_{i,i}=0$ for all $i=1,\ldots,R$.
  
The graph Laplacian $\mathcal L^{(s)}= D^{(s)} - A^{(s)}$ is a routinely used one-to-one transform of $A^{(s)}$, where $D^{(s)}$ is a diagonal matrix $D^{(s)}_{ii}=\sum_{j=1}^R A_{i,j}^{(s)}$. Compared to the adjacency matrix, the Laplacian enjoys a few appealing properties, namely: (i) $\mathcal L^{(s)}$ is always positive semidefinite, (ii) the number of zero eigenvalues equals the number of disjoint component sub-graphs (each known as a community); (iii) the  smallest non-zero eigenvalues quantify the  connectivity (normalized graph cut) in each component sub-graph. Because of these properties, we can quantify the difference between two graphs via the geodesic distance in the interior of the positive definite cone \citep{lim2019geometric}. For two positive semidefinite matrices $X$ and $Y$ of equal size, $\text{dist}(X,Y) = \lim_{\eta\to 0_+}\Bigl(\sum_{j=1}^R \log^2\bigl[\xi_j\{(X+I\eta)^{-1}(Y+I\eta)\}\bigr]\Bigr)^{1/2}$, where $\xi_j(\cdot)$ is the $j$-th eigenvalue.

\begin{figure}[H]
	\centering
	\begin{subfigure}[t]{0.45\textwidth}
		\centering
		\includegraphics[width=\linewidth]{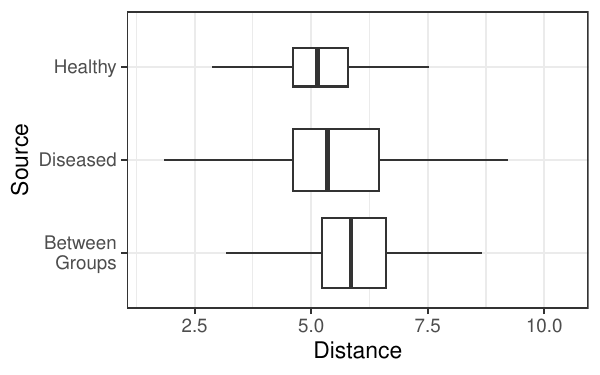}
		\caption{Boxplot of pairwise distances of the observed Laplacian matrices.}
	\end{subfigure}
	\begin{subfigure}[t]{0.45\textwidth}
		\centering
		\includegraphics[width=\linewidth]{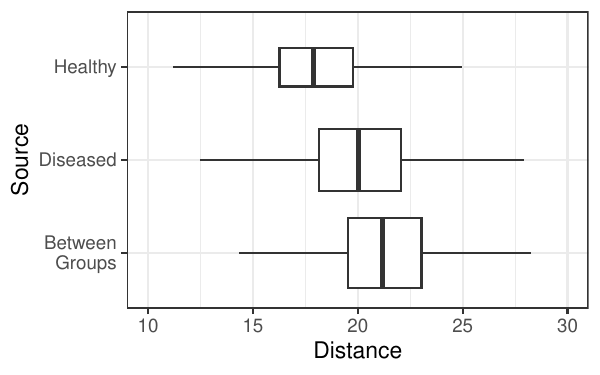}
		\caption{Boxplot of pairwise distances of the low-rank smoothed Laplacian matrices using different $\lambda_s$ for each subject.}
		\label{fig:pair_dist}
	\end{subfigure}
	\caption{Boxplots of the pairwise distances among the observed Laplacian matrices, and that among the smoothed Laplacian matrices.}
	\label{fig:HFCG_boxplots}
\end{figure}

Figure \ref{fig:HFCG_boxplots}(a) plots the pairwise distances between the observed $\mathcal L^{(s)}$ using three boxplots corresponding to subjects within the diseased group, subjects within the healthy group, and those between the two groups. While we see that the within-group distances have slightly smaller means than the between-group distances, there is significant overlap among the three boxplots. This is understandable since each observed $\mathcal L^{(s)}$ is of rank $(R-1)$, corresponding to having no disjoint components (i.e. only one community), motivating a reduced-rank modeling approach.

We therefore consider each $\mathcal L^{(s)}$ as generated near a manifold $\mathcal M^{(s)}$, each with a Gaussian-type density proportional to $\exp\left\{ -{  \text{dist}(\mathcal L^{(s)}, \mathcal M^{(s)})^2}/{2\sigma^2} \right\}$. This manifold $\mathcal M^{(s)}$ is given by the intersection between the space of Laplacians and the nuclear norm ball of radius $r^{(s)}>0$:
\(
\mathcal M^{(s)} = \biggl\{ \|\zeta\|_* \le r^{(s)}, \zeta \in \mathbb{R}^{R\times R} \mid \zeta_{i,i}=-\sum_{j:j\neq i}\zeta_{i,j} , \zeta_{i,j}=\zeta_{j,i} \leq 0 \text{ for } i \neq j \biggr\}.
\)
Here we denote $\|\zeta\|_*$ the nuclear norm of matrix $\zeta$, given by the sum of singular values of $\zeta$. Akin to how the $\ell_1$ ball promotes sparsity, the boundary of the nuclear norm ball coincides with matrices of low rank. Since we do not know $r^{(s)}$ (hence do not know $\mathcal M^{(s)}$ completely), we want to quantify the uncertainty in the bridged posterior framework.

The distance to $\mathcal M^{(s)}$ can be computed by first solving for the projection:
\begin{equation}\label{eq:projection}
\begin{aligned}
 &     Z^{(s)} = {\arg\min}_{\zeta}\ \frac{1}{2}\|\mathcal L^{(s)}-\zeta\|_F^2 + \tilde\lambda_s (\|\zeta\|_* -r^{(s)})\\
 &\text{subject to }  \zeta\in \mathbb{R}^{n\times n},  \zeta_{i,i}=-\sum_{j:j\neq i}\zeta_{i,j} , \zeta_{i,j}=\zeta_{j,i} \leq 0 \text{ for } i \neq j,
\end{aligned}
\end{equation}
where $\tilde\lambda_s\ge 0$ is a Lagrange multiplier, and $\|Z^{(s)}\|_* \le r^{(s)}$; then we have 
$\text{dist}(\mathcal L^{(s)}, \mathcal M^{(s)})= \|\mathcal L^{(s)}-Z^{(s)}\|_F$. Although we do not know $r^{(s)}$ in advance, we know if we were given the value of $\tilde\lambda_s>0$, then $\tilde Z^{(s)}={\arg\min}_{\zeta}\ \frac{1}{2}\|\mathcal L^{(s)}-\zeta\|_F^2 + \tilde\lambda_s \|\zeta\|_*$ would be the same solution of \eqref{eq:projection} given $r^{(s)}=\|\tilde Z^{(s)}\|_*$ (provided $\mathcal L^{(s)}$ is outside $\mathcal M^{(s)}$). Therefore, assigning a prior to $\tilde\lambda_s>0$ is equivalent to assigning a prior on $r^{(s)}$.

The question boils down to how to meaningfully model the collection of $\tilde\lambda_s$.

Due to the heterogeneity of $\mathcal L^{(s)}$, the same value of $\tilde \lambda_s=\tilde \lambda_{s'}$ may yield quite different $Z^{(s)}$ and $Z^{(s')}$. As a result, for the purpose of data harmonization, instead of assigning  independent priors or setting equal values for $\tilde \lambda_s$, we assign a dependent likelihood based on the pairwise distances among $Z^{(s)}$:
\(
L \Bigl[ \{\mathcal {L}^{(s)}, Z^{(s)}\}_{s=1}^S ; \{\tilde\lambda_s\}_{s=1}^S, \sigma^2, \tau \Bigr] & \propto 
\biggl[
\prod_{s=1}^S (\sigma^2)^{-1/2}\exp\biggl\{ -\frac{\|\mathcal L^{(s)}-Z^{(s)}\|_F^2}{2\sigma^2} \biggr\}  \frac{\tilde\lambda_s}{\sigma^2} \exp\biggl\{ -\frac{\tilde\lambda_s\|Z^{(s)}\|_*}{\sigma^2} \biggr\}       \biggr]\\
& \qquad \times \biggl[  \prod_{s=1}^{S} \tau ^{-1/2} \exp\biggl\{ -\frac{\sum_{k:k\neq s}\text{dist}^2\bigl(Z^{(k)}, Z^{(s)}\bigr)/(S-1)}{2\tau} \biggr\}  \biggr].
\)
The second line is a pairwise kernel via the average total squared geodesic distance between each $Z^{(s)}$ and other $Z^{(s')}$, so that it borrows information across subjects to reduce the heterogeneity.  We clarify that group information is not used above; hence it can serve as a data harmonization tool, even in the absence of group labels.

To calculate $Z^{(s)}$ at different $\tilde\lambda_s$, we use the alternating direction method of multipliers (ADMM) algorithm (details are provided in \cref{sec:admm}). To facilitate computation, for each $\tilde \lambda_s$, we assign a discrete uniform equally spread over $10$ values in $(0,5]$, so that the possible values of $Z^{(s)}$ as well as their associated pairwise geodesic distances can be precomputed before running MCMC. We specify an Inverse-Gamma$(2,1)$ prior for $\sigma^2$ and Inverse-Gamma$(2,1)$ prior for $\tau$.  Running MCMC for $10,000$ iterations takes {46.5} minutes on a 12-core laptop; the first $2,000$ samples are discarded as burn-in. 

Using the smoothed Laplacian $Z^{(s)}$, we calculate posterior mean of the  the distance matrix $\{ \text{dist}\bigl(Z^{(s)}, Z^{(s')}\bigr)\}_{\text{all }s,s'}$ and show re-calculated boxplots of geodesic distances in Figure \ref{fig:HFCG_boxplots}(b). Clearly, between the low-rank smoothed $Z^{(s)}$, the healthy group now has much lower pairwise distances than the diseased group; and the diseased group has slightly lower pairwise distances compared to the between-group. We compute the Kolmogorov–Smirnov (KS) statistical metric between the empirical distribution of geodesic distances.  When we switch from using raw $\mathcal L^{(s)}$ to smoothed $Z^{(s)}$, the KS metric between the diseased and healthy increases from {0.143} to {0.299}.

\begin{figure}[H]
	\begin{subfigure}[t]{0.45\textwidth}
		\centering
		\includegraphics[width=1\linewidth]{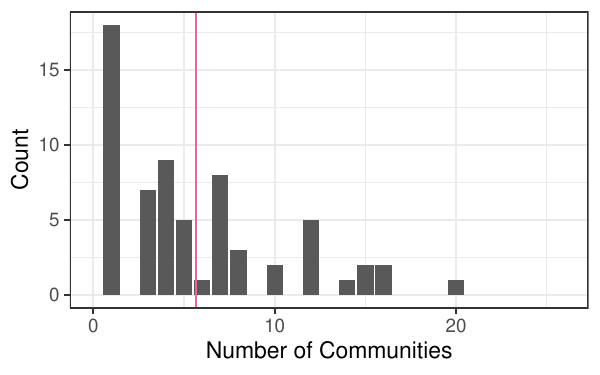}
		\caption{Healthy group (64 subjects).}
	\end{subfigure}\qquad
	\begin{subfigure}[t]{0.45\textwidth}
		\centering
		\includegraphics[width=1\linewidth]{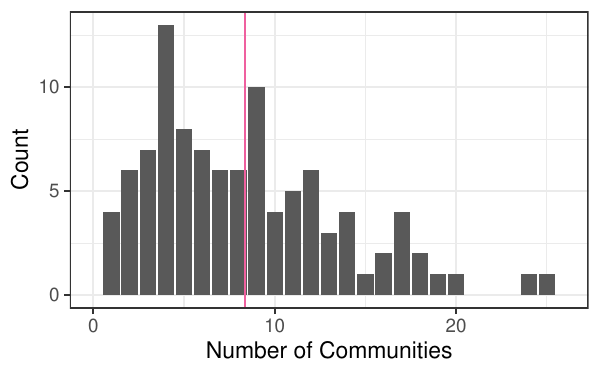}
		\caption{Diseased group  (102 subjects).} 
	\end{subfigure}
	\centering
	\caption{The barplots on the number of communities in $Z^{(s)}$ at each subject's posterior mean $\lambda_s$. The vertical line is the mean of the number of communities over all subjects.
		\label{fig:hist}}
\end{figure}

By calculating the number of zero eigenvalues of the $Z^{(s)}$, we find $K^{(s)}$ as the number of communities for each subject. Figure \ref{fig:hist} shows histograms of $K^{(s)}$ evaluated at each subject's posterior mean $\tilde\lambda^s$. The average number of communities for the healthy subjects is $5.77$ while it is $8.41$ for the diseased subjects. This is consistent with the known fact that a diseased brain tends to be more fragmented than a healthy one, due to the disruptions caused by Alzheimer's disease.

\begin{figure}[H]
	\begin{minipage}[c]{0.6\textwidth}
		\begin{subfigure}[t]{0.45\textwidth}
			\centering
			\includegraphics[width=1\linewidth,trim={3cm 3cm 0cm 3cm},clip]{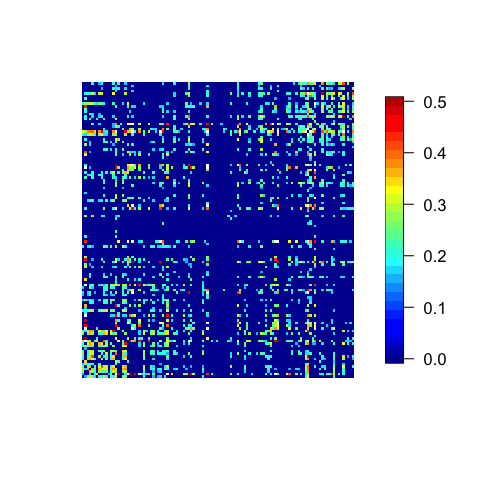}
			\caption*{Healthy subject 1.}
		\end{subfigure}
		\begin{subfigure}[t]{0.45\textwidth}
			\centering
			\includegraphics[width=1\linewidth,trim={3cm 3cm 0cm 3cm},clip]{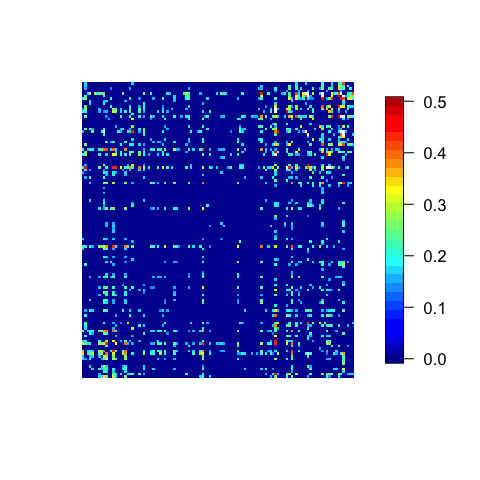}
			\caption*{Healthy subject 2.}
		\end{subfigure}\\
		\begin{subfigure}[t]{0.45\textwidth}
			\centering
			\includegraphics[width=1\linewidth,trim={3cm 3cm 0cm 3cm},clip]{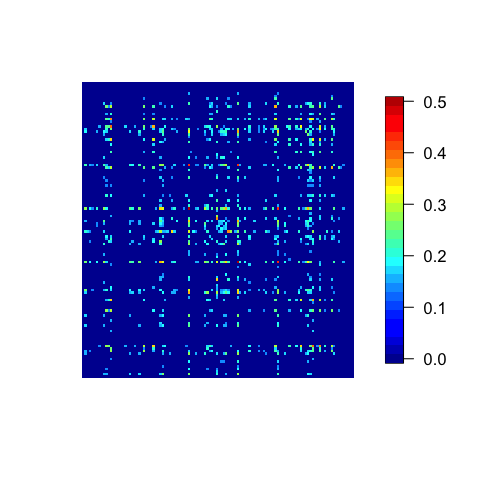}
			\caption*{Diseased subject 1.}
		\end{subfigure}
		\begin{subfigure}[t]{0.45\textwidth}
			\centering
			\includegraphics[width=1\linewidth,trim={3cm 3cm 0cm 3cm},clip]{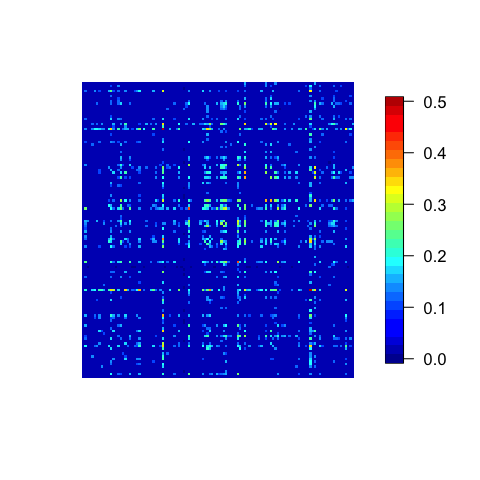}
			\caption*{Diseased subject 2.}
		\end{subfigure}\\
		(a) Smoothed adjacency matrices for four subjects.
	\end{minipage}
	\begin{minipage}[c]{0.4\textwidth}
		\centering
		\begin{subfigure}[t]{1\textwidth}
			\includegraphics[width=1\linewidth]{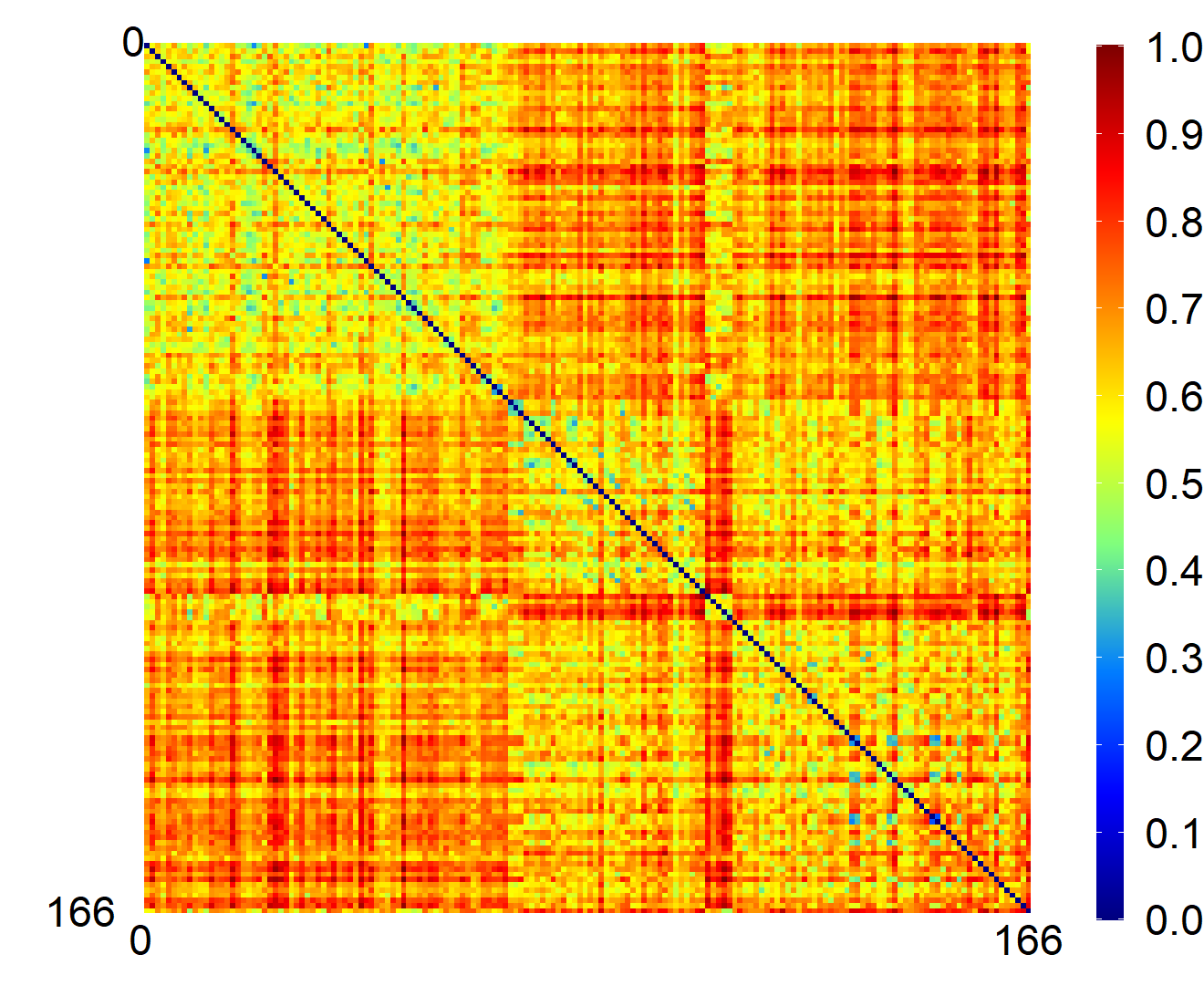}
			\caption*{Top-left $64\times 64$: healthy subjects, bottom-right $102\times 102$: diseased subjects.}
		\end{subfigure} \vspace*{.6cm} \\ 
		(b) Posterior mean of the pairwise geodesic distance among the smoothed Laplacians.
	\end{minipage}
	\caption{Illustration of the smoothed graph estimates. Panel(a) plots the smoothed adjacency matrices for four subjects, based on $\tilde A^{(s)} = -Z^{(s)}$  (the diagonal elements are masked) with $Z^{(s)}$ obtained at the posterior mean of $\lambda^{(s)}$ for each $s$, Panel(b) shows the posterior mean matrix of the geodesic distances between $Z^{(1)},\ldots, Z^{(S)}$.}
	\label{fig:illustration_networks}
\end{figure}

Figure \ref{fig:illustration_networks} shows the smoothed adjacency matrices for two subjects chosen from the healthy group, and two from the diseased group, and the posterior mean of the pairwise geodesic distances. To validate the result, we further apply spectral clustering on the pairwise distance matrix, and cluster the subjects into two groups. Based on the posterior mean distance matrix among $Z^{(s)}$, $96.9 \%$ of the subjects in the healthy group are correctly grouped together, and $89.2\%$ for the diseased group. Using the distance matrix among raw $\mathcal L^{(s)}$, these numbers are $87.5\%$ and $89.2\%$.

\begin{remark}
    Using bridged posterior has a clear computational advantage over the canonical model with the smoothed matrices being completely random. It is conceivably difficult to sample from the canonical posterior, as it would involve drawing $S$ many high-dimensional matrices. From a data harmonization perspective, practitioners typically want to produce a single matrix per subject; hence, the bridged posterior (using projection) provides a straightforward solution.
\end{remark}

\section{Discussion}

In this article, we present an approach for using optimization as a modeling tool to form a class of augmented likelihoods. 
These likelihoods enjoy a generative interpretation, with the use of latent variables $z$ and constraints via the conditional distribution of $y$ given $z$. Hence, they are amenable to the inference in the Bayesian framework, and in turn allow uncertainty quantification. We demonstrate several computational and modeling advantages over related Gibbs posterior alternatives in the literature.

In our present examples, we have focused on well-behaved loss functions with unique optima, which can be obtained efficiently with high numerical accuracy. Moving beyond this relatively clean setting merits future study as extensions and generalizations of the bridged posterior. Many problems, such as those with non-convex objectives functions, entail losses featuring multiple local optima \citep{zhang2020}. The solution $z$ returned by the algorithm at convergence may depend on the choice of initialization $\tilde z$ in these cases, where our approach has made use of a well-defined solution as input in a hierarchy. One generalization that may be fruitful is to assign a probability distribution over $\tilde z$, enabling us to view the optimization procedure as an algorithm mapping to another distribution for $z$. Second, many popular optimization algorithms, including stochastic variants of schemes such as gradient descent and early stopping, may produce approximately optimal solutions. In such cases, it may be more appropriate to model $\Pi(z\mid y,\lambda)$ to carry some uncertainty reflecting numerical errors or stopping criteria, in place of the point mass used in our current formulation. It is interesting to explore further connections to areas including Bayesian probabilistic numerical methodology \citep{cockayne2019bayesian} and optimization-based frequentist confidence intervals for constrained problems \citep{batlle2023}. 

Empirically we observe a faster mixing rate of the Markov chain for the bridged posterior, compared to the chain targeting the joint posterior under canonical full likelihood. A related theory can be found in \cite{liu1994collapsed}, who proved that the Markov chain targeting the integrated posterior can have a faster mixing rate than the chain targeting the associated joint posterior. On the other hand, this intuition is not directly extendable to our case, as the bridged posterior and the integrated posterior are often not the same. Therefore, a rigorous analysis on comparing the mixing rates of the two Markov chains remains illusive at this point, and can be pursued as a future work.

\bibliographystyle{plainnat}
\bibliography{ref}

\appendix

\section{Proofs}
\label{sec:proofs}

The proof of \cref{thm:BvM_parametric} uses a theorem from \cite{miller2021asymptotic}. We provide the complete statement of that theorem as following.

\renewcommand{\theenumi}{\arabic{enumi}}
\renewcommand{\labelenumi}{\theenumi.\ }

\begin{theorem}{\citet[Theorem~5]{miller2021asymptotic}}\label{thm:miller}
	Let $\Theta \subseteq \mathbb{R}^{D}$. Let $E \subseteq \Theta$ be open (in $\mathbb{R}^{D}$) and bounded. Fix $\theta_{0} \in E$ and let $\pi: \Theta \rightarrow \mathbb{R}$ be a probability density with respect to Lebesgue measure such that $\pi$ is continuous at $\theta_{0}$ and $\pi(\theta_{0})>0$. Let $f_{n}: \Theta \rightarrow \mathbb{R}$ have continuous third derivatives on $E$. Suppose $f_{n} \rightarrow f$ pointwise for some $f: \Theta \rightarrow \mathbb{R}, f^{\prime \prime}(\theta_{0})$ is positive definite, and $(f_{n}^{\prime \prime \prime})$ is uniformly bounded on E. If either of the following two assumptions is satisfied:
	\begin{enumerate}
		\item $f(\theta)>f(\theta_{0})$ for all $\theta \in K \backslash\{\theta_{0}\}$ and $\liminf _{n} \inf _{\theta \in \Theta \backslash K} f_{n}(\theta)>f(\theta_{0})$ for some compact $K \subseteq E$ with $\theta_{0}$ in the interior of $K$, or
		\item each $f_{n}$ is convex and $f^{\prime}(\theta_{0})=0$,
	\end{enumerate}
	then there is a sequence $\theta_{n} \rightarrow \theta_{0}$ such that $f_{n}^{\prime}(\theta_{n})=0$ for all $n$ sufficiently large, $f_{n}(\theta_{n}) \rightarrow$ $f(\theta_{0})$, defining $m_{n}=\int_{\mathbb{R}^{D}} \exp (-n f_{n}(\theta)) \pi(\theta)\, \textup{d} \theta$  and  $\pi_{n}(\theta)=\exp (-n f_{n}(\theta)) \pi(\theta) / m_{n}$, we have
	$\int_{B_{\varepsilon}(\theta_{0})} \pi_{n}(\theta)\, \textup{d} \theta \xrightarrow [n \to \infty]{} 1 \text { for all } \varepsilon>0$,
	that is, $\pi_{n}$ concentrates at $\theta_{0}$, and letting $q_{n}$ be the density of $\sqrt{n}(\theta-\theta_{n})$ when $\theta \sim \pi_{n}$, we have $\int_{\mathbb{R}^{D}}\big|q_{n}(x)-\mathcal{N}\bigl(x \mid 0, H_{0}^{-1}\bigr)\big|\,\textup{d} x \xrightarrow [n \to \infty]{} 0$, that is, $q_{n}$ converges to $\mathcal{N}(0, H_{0}^{-1})$ in total variation, where $H_{0}=f^{\prime \prime}(\theta_{0})$. Further, $2 \Rightarrow 1$ under the assumptions of the theorem.
\end{theorem}

\begin{proof}[of Theorem \ref{thm:BvM_parametric}]
	We show the following properties in the sense
	of $a.s.[y_{1:n}]$.
	\begin{enumerate}[leftmargin=0.4in, itemsep=0pt, topsep=0pt]
		\item $\hat{l}_n$ has continuous third derivatives on $E$. Since $l_n$ has continuous third derivatives on $E\times \hat{z}_n(E) $ and $\hat{z}_n$ has continuous third derivative by Assumption (A1), we have
		\(
		\hat{l}_n'(\lambda) & = \frac{\partial l_n(\lambda, \zeta)}{\partial\lambda}\bigg|_{ \zeta =\hat{z}_n(\lambda)} + \frac{\partial l_n(\lambda, \zeta)}{\partial  \zeta }\bigg|_{ \zeta =\hat{z}_n(\lambda)} \hat{z}_n'(\lambda),\\
		\hat{l}_n''(\lambda) & = \frac{\partial^2 l_n(\lambda, \zeta)}{\partial\lambda^2} \bigg|_{ \zeta =\hat{z}_n(\lambda)} +2 \frac{\partial^2 l_n(\lambda, \zeta)}{\partial\lambda\partial \zeta } \bigg|_{ \zeta =\hat{z}_n(\lambda)} \hat{z}_n'(\lambda) \\
		& \qquad +[\hat{z}_n'(\lambda)]^\top \frac{\partial^2 l_n(\lambda, \zeta)}{\partial \zeta ^2} \bigg|_{ \zeta =\hat{z}_n(\lambda)} \hat{z}_n'(\lambda)+\frac{\partial l_n(\lambda, \zeta)}{\partial  \zeta }\bigg|_{ \zeta =\hat{z}_n(\lambda)} \hat{z}_n''(\lambda).
		\)
		Then it is not hard to see that the property is satisfied.
		\item $\hat{l}_n\to \hat{l}_*$ pointwise on $\Theta$ and $-\hat{l}_*''(\lambda_0)$ is positive definite. The first part is by Assumption (A2), $l_n \to l_*$
		and $\hat{z}_n\to \hat{z}_*$. To show $-\hat{l}_*''(\lambda_0)$ is positive definite,
		we have
		\(
		\hat{l}_*''(\lambda_0) & = \frac{\partial^2 l_*(\lambda_0, \zeta)}{\partial\lambda^2} \bigg|_{ \zeta =\hat{z}_*(\lambda_0)} +2 \frac{\partial^2 l_*(\lambda_0, \zeta)}{\partial\lambda\partial \zeta } \bigg|_{ \zeta =\hat{z}_*(\lambda_0)} \hat{z}_*'(\lambda_0) \\
		& \qquad + [\hat{z}_*'(\lambda_0)]^\top \frac{\partial^2 l_*(\lambda_0, \zeta)}{\partial \zeta ^2} \bigg|_{ \zeta =\hat{z}_*(\lambda_0)} \hat{z}_*'(\lambda_0)+\frac{\partial l_*(\lambda_0, \zeta)}{\partial  \zeta }\bigg|_{ \zeta =\hat{z}_*(\lambda_0)} \hat{z}_*''(\lambda_0) \\
		& = \begin{bmatrix} I_{d} & [\hat{z}_*'(\lambda_0)]^\top  \end{bmatrix} l_*''(\lambda_0, \hat{z}_*(\lambda_0))\begin{bmatrix} I_{d} \\ \hat{z}_*'(\lambda_0) \end{bmatrix}
		\)
		where the second equation is using $\frac{\partial l_*(\lambda_0, \zeta)}{\partial  \zeta }\big|_{ \zeta =\hat{z}_*(\lambda_0)}=0$ to cancel out the last term. Note that $-l_*''(\lambda_0, \hat{z}_*(\lambda_0))$ is positive definite by Assumption (A3).
		\item $\hat{l}_n'''$ is uniformly bounded on $E$. Since $l_n'''$ and $\hat{z}_n'''$ are uniformly bounded, by the theorem 7 of \citet{miller2021asymptotic}, $l_n',\ l_n'', \hat{z}_n'$
		and $\hat{z}_n''$ are all uniformly bounded. Hence, $\hat{l}_n'''$ is uniformly bounded by the expansion of $\hat{l}_n'''$:
		\(
		\hat{l}_n'''(\lambda) ={} & \frac{\partial^3 l_n(\lambda, \zeta)}{\partial\lambda^3} \bigg|_{ \zeta =\hat{z}_n(\lambda)} + 3 \frac{\partial^3 l_n(\lambda, \zeta)}{\partial\lambda^2\partial \zeta } \bigg|_{ \zeta =\hat{z}_n(\lambda)} \hat{z}_n'(\lambda) \\
		& + 3 [\hat{z}_n'(\lambda)]^\top \frac{\partial^3 l_n(\lambda, \zeta)}{\partial \lambda \partial \zeta ^2} \bigg|_{ \zeta =\hat{z}_n(\lambda)} \hat{z}_n'(\lambda) + [\hat{z}_n'(\lambda)]^\top \frac{\partial^3 l_n(\lambda, \zeta)}{\partial \zeta ^3} \bigg|_{ \zeta =\hat{z}_n(\lambda)} [\hat{z}_n'(\lambda)]^2 \\
		& + 3 [\hat{z}_n'(\lambda)]^\top \frac{\partial^2 l_n(\lambda, \zeta)}{\partial \zeta ^2} \bigg|_{ \zeta =\hat{z}_n(\lambda)} \hat{z}_n''(\lambda) + 3\frac{\partial^2 l_n(\lambda, \zeta)}{\partial \lambda \partial  \zeta }\bigg|_{ \zeta =\hat{z}_n(\lambda)} \hat{z}_n''(\lambda) \\
            & + \frac{\partial l_n(\lambda, \zeta)}{\partial  \zeta }\bigg|_{ \zeta =\hat{z}_n(\lambda)} \hat{z}_n'''(\lambda).
		\)
		\item By Assumption (A4), for some compact $K \subseteq E$ with $\lambda_{0}$ in the interior of $K$, $\hat{l}_*(\lambda) < \hat{l}_*(\lambda_{0})$ for all $\lambda \in K \backslash\{\lambda_{0}\}$
		and\ $\limsup_{n} \sup_{\lambda \in \Theta \backslash K} \hat{l}_n(\lambda) < \hat{l}_*(\lambda_{0})$.
	\end{enumerate}
	By Theorem \ref{thm:miller} with $f_n=-\hat{l}_n$ and $f=-\hat{l}_*$, the above 1--4 complete the proof.
\end{proof}

\begin{proof}[of \cref{coro:smallvar}]
	We use the delta method to find the asymptotic distribution of $\sqrt{n}$-adjusted $\hat{z}_n(\lambda)$. Since convergence in total variation implies convergence in distribution (weak convergence), the random vector $\sqrt{n}(\lambda-\lambda_{n}) \rightsquigarrow \text{N}(0,H_0^{-1})$. Using delta method, we can prove $\sqrt{n}\{\hat{z}_n(\lambda)-\hat{z}_n(\lambda_{n})\} \rightsquigarrow \text{N}(0, H_z^{-1})$ where $H_z^{-1} = \hat{z}'_*(\lambda_0)H_0^{-1} \hat{z}'_*(\lambda_0)^\top$.

	When $g_n(\zeta, y_{1:n};\lambda) = -L(y_{1:n},\zeta; \lambda)$ with that the bridged posterior coincides with the profile likelihood, the asymptotic variance above can be represented by the second derivatives of $l_*$. We first show that $-H_0= \hat{l}''_*(\lambda_0)= l_{*,\lambda_0 \lambda_0} -l_{*,\lambda_0 \zeta_0}l_{*,\zeta_0\zeta_0}^{-1} l_{*,\zeta_0\lambda_0}$, 
    where $l_{*,\lambda_0\lambda_0}, l_{*,\zeta_0\zeta_0}, l_{*,\zeta_0\lambda_0}, l_{*,\lambda_0 \zeta_0}$ are respectively the second partial derivatives
	${\partial^2 l_*(\lambda, \zeta)}/{\partial \lambda^2}$, ${\partial^2 l_*(\lambda, \zeta)}/{\partial \zeta^2}$, ${\partial^2 l_*(\lambda, \zeta)}/{\partial \zeta\partial \lambda }$, ${\partial^2 l_*(\lambda, \zeta)}/{\partial \lambda \partial \zeta}$ evaluating at $\lambda=\lambda_0,\zeta=\zeta_0=\hat{z}_*(\lambda_0)$. 
	
	Since in this case $\frac{\partial l_n(\lambda, \zeta)}{\partial \zeta}|_{\zeta=\hat{z}_n(\lambda)}=0$, we have
	\(0=\frac{\partial }{\partial \lambda} \biggl\{ \frac{\partial l_n(\lambda, \zeta)}{\partial \zeta} \bigg|_{\zeta=\hat{z}_n(\lambda)}\biggr\}
	= \frac{\partial^2 l_n(\lambda, \zeta)}{\partial \zeta\partial \lambda }\bigg|_{\zeta=\hat{z}_n(\lambda)} + \frac{\partial^2 l_n(\lambda, \zeta)}{\partial \zeta^2}\bigg|_{\zeta=\hat{z}_n(\lambda)} \hat{z}_n'(\lambda).\)
	Hence
	\(
	\hat{z}_n'(\lambda) = - \biggl\{ \frac{\partial^2 l_n(\lambda, \zeta)}{\partial \zeta^2}\bigg|_{\zeta=\hat{z}_n(\lambda)} \biggr\}^{-1}  \frac{\partial^2 l_n(\lambda, \zeta)}{\partial \zeta\partial \lambda }\bigg|_{\zeta=\hat{z}_n(\lambda)}.
	\)
	By \citet[Theorem 7]{miller2021asymptotic}, we have $l_n'' \to l_*''$. Letting $n\to\infty$ and $\lambda=\lambda_0$, we have $\hat{z}'_*(\lambda_0) = -l_{*,\zeta_0\zeta_0}^{-1} l_{*,\zeta_0\lambda_0}$.
	Now
	\(
	\hat{l}''_n(\lambda)  = \frac{\partial }{\partial \lambda} \biggl\{\frac{\partial l_n(\lambda, \zeta)}{\partial \lambda}\bigg|_{\zeta=\hat{z}_n(\lambda)}\biggr\}
	= \frac{\partial^2 l_n(\lambda, \zeta)}{\partial \lambda \partial \zeta }\bigg|_{\zeta=\hat{z}_n(\lambda)} \hat{z}_n'(\lambda) + \frac{\partial^2 l_n(\lambda, \zeta)}{\partial \lambda^2}\bigg|_{\zeta=\hat{z}_n(\lambda)} .
	\)
	Letting $n\to\infty$ and $\lambda=\lambda_0$, we have the result $\hat{l}''_*(\lambda_0)= l_{*,\lambda_0 \lambda_0} -l_{*,\lambda_0 \zeta_0}l_{*,\zeta_0\zeta_0}^{-1} l_{*,\zeta_0\lambda_0}$.
	
	By Assumption (A3), both $-l_{*,\lambda_0\lambda_0}$ and $-l_{*,\zeta_0\zeta_0}$ are positive definite.
	The asymptotic variance of $\sqrt{n}\{\hat{z}_n(\lambda)-\hat{z}_n(\lambda_{n})\}$ is thus\\ $\hat{z}'_*(\lambda_0)H_0^{-1} \hat{z}'_*(\lambda_0)^\top = -l_{*,\zeta_0\zeta_0}^{-1} l_{*,\zeta_0\lambda_0} (l_{*,\lambda_0 \lambda_0} -l_{*,\lambda_0 \zeta_0}l_{*,\zeta_0\zeta_0}^{-1} l_{*,\zeta_0\lambda_0})^{-1} l_{*,\lambda_0 \zeta_0} l_{*,\zeta_0\zeta_0}^{-1} $. 
	
	If we treat the latent variable $\zeta$ as non-deterministic in the likelihood $L(y,\zeta; \lambda)$ with some prior, then \cite{miller2021asymptotic} proves $\sqrt{n}\bigl([ \lambda \ \zeta] ^\top - [ \lambda_n \ \zeta_n ] ^\top
	\bigr) \rightsquigarrow N(0,\tilde{H}_0^{-1})$ for some sequences $\lambda_n$ and $\zeta_n$ when $(\lambda,\zeta)\sim \Pi(\lambda,\zeta\mid y)$, where $\tilde{H}_0 = -l''_*(\lambda_0, \zeta_0)$ and $\zeta_0=\hat{z}_*(\lambda_0)$ is a fixed ground-truth of $\zeta$. Marginally, the asymptotic variance of $\sqrt{n}(\zeta-\zeta_n)$ is the $\zeta$-block of $\tilde{H}_0^{-1}$, which is equal to
	\(
	-(l_{*, \zeta_0\zeta_0} - l_{*,\zeta_0\lambda_0 }l_{*,\lambda_0\lambda_0}^{-1} l_{*,\lambda_0 \zeta_0})^{-1} 
	= -l_{*, \zeta_0\zeta_0}^{-1} - l_{*, \zeta_0\zeta_0}^{-1} l_{*,\zeta_0\lambda_0} (l_{*,\lambda_0 \lambda_0} -l_{*,\lambda_0 \zeta_0}l_{*,\zeta_0\zeta_0}^{-1} l_{*,\zeta_0\lambda_0})^{-1} l_{*,\lambda_0 \zeta_0} l_{*,\zeta_0\zeta_0}^{-1}.
	\)
	Since $-l_{*, \zeta_0\zeta_0}^{-1}$ is positive definite, the asymptotic variance of the $j$-th element of $\sqrt{n}(\zeta-\zeta_n)$ is strictly greater than the one of the $j$-th element of $\sqrt{n}\{\hat{z}_n(\lambda)-\hat{z}_n(\lambda_{n})\}$.
\end{proof}

\begin{proof}[of Lemma \ref{lemma:LAN}]
	By the definition of the profile likelihood, we have $\hat{l}_n(\lambda) = l_n(\lambda, \hat{z}_{\lambda}) \geq l_n\{ \lambda, \tilde \zeta_\lambda(\lambda_0, \hat{z}_{\lambda_0}) \}$, so
	\begin{align}
		\hat{l}_{n}(\lambda)- \hat{l}_{n}(\lambda_0)
		& =  l_n(\lambda, \hat{z}_{\lambda})- l_n(\lambda_0, \hat{z}_{\lambda_0}) \notag\\
		& \geq l_n\{ \lambda, \tilde{\zeta}_\lambda(\lambda_0, \hat{z}_{\lambda_0}) \} - l_n\{ \lambda_0, \tilde{\zeta}_{\lambda_0}(\lambda_0, \hat{z}_{\lambda_0}) \} \notag\\
		& = \tilde{l}_n(\lambda, \lambda_0, \hat{z}_{\lambda_0}) - \tilde{l}_n(\lambda_0, \lambda_0, \hat{z}_{\lambda_0}) \label{eq:lo}
	\end{align}
	by using $\tilde{\zeta}_\lambda(\lambda, \zeta) = \zeta$ for the second term. Similarly, $\hat{l}_n(\lambda_0) = l_n(\lambda_0, \hat{z}_{\lambda_0}) \geq l_n\{\lambda_0, \tilde \zeta_{\lambda_0}(\lambda, \hat{z}_\lambda)\}$, and hence
	\begin{align}
		\hat{l}_{n}(\lambda)- \hat{l}_{n}(\lambda_0)
		& =  l_n(\lambda, \hat{z}_{\lambda})- l_n(\lambda_0, \hat{z}_{\lambda_0}) \notag\\
		& \leq l_n\{(\lambda, \tilde{\zeta}_\lambda({\lambda}, \hat{z}_{\lambda})\} - l_n\{\lambda_0, \tilde{\zeta}_{\lambda_0}(\lambda, \hat{z}_{\lambda})\} \notag \\
		& = \tilde{l}_n(\lambda, {\lambda}, \hat{z}_{\lambda})- \tilde{l}_n(\lambda_0, \lambda, \hat{z}_{\lambda}) \label{eq:up}
	\end{align}
	by using $\tilde{\zeta}_\lambda(\lambda, \zeta) = \zeta$ for the first term. Both \eqref{eq:lo} and \eqref{eq:up} are differences between function $l_n$ evaluating at $\lambda$ and the one at $\lambda_0$ while keeping the other arguments unchanged. They have the Taylor expansion of the function $\tilde{l}_n$ with respect to the first argument,
	\[
	(\lambda - \lambda_0)^\top \frac{\partial \tilde{l}_n(t, \psi, \hat{z}_\psi)}{\partial t}\bigg|_{t=\lambda_0} + \frac{1}{2}(\lambda - \lambda_0)^\top \frac{\partial^2 \tilde{l}_n(t, \psi, \hat{z}_\psi)}{\partial t^2}\bigg|_{t=\tilde{t}} (\lambda - \lambda_0)
	\label{eq:taylor}
	\]
	where $\tilde{t}$ is somewhere between $\lambda$ and $\lambda_0$, and $\psi$ can be $\lambda$ or $\lambda_0$. By the assumption 1 in (B1), the second term is equal to $-  (\lambda - \lambda_0)^\top H_0(\lambda-\lambda_0) / 2 + o_{P_{\lambda_0,\zeta_0}}(1)(\|\lambda-\lambda_0\|^2)$. By the assumption 2 in (B1), the first term is equal to
	\(
	(\lambda-\lambda_0)^\top h_n + (\lambda-\lambda_0)^\top \mathbb{E}_{\lambda_0,\zeta_0} \frac{\partial \tilde{l}_n(t, \lambda, \hat{z}_\lambda)}{\partial t}\bigg|_{t=\lambda_0} + o_{P_{\lambda_0,\zeta_0}}(1) \bigl( \|\lambda-\lambda_0\| n^{-1/2} \bigr).
	\)
	Combining with \eqref{eq:expect_score} and $\|\lambda-\lambda_0\| n^{-1/2}\leq \bigl(\|\lambda-\lambda_0\| + n^{-1/2} \bigr)^2$, the first term of \eqref{eq:taylor} becomes
	$(\lambda-\lambda_0)^\top h_n + o_{P_{\lambda_0,\zeta_0}}(1)\bigl\{\bigl(\|\lambda-\lambda_0\| + n^{-1/2} \bigr)^2\bigr\}$, and hence \eqref{eq:lan} is proved.
\end{proof}

\begin{proof}[of \cref{thm:BvM_semi}]
	We have $\pi_n(\lambda) = \exp\{n\hat{l}_n(\lambda)\} \pi_0(\lambda) / m_n$ where $m_n = \int_{\mathbb{R}^d} \exp\{ n\hat{l}_n(\lambda)\}\pi_0(\lambda) \,\textup{d}\lambda$, and
	$q_n(x) = \pi_n(\lambda_n+x/\sqrt{n}) n^{-d/2}$. Let
	\(g_n(x) =  q_n(x)\exp\{ -n\hat{l}_n(\lambda_n)\} n^{d/2}m_n = \exp[ n\{\hat{l}_n(\lambda_n + x/\sqrt{n}) - \hat{l}_n(\lambda_n)\} ] \pi_0(\lambda_n+x/\sqrt{n})
	\)
	and define $g_0(x) = \exp\{ -x^\top H_0 x / 2\}\pi_0(\lambda_0)$.
	We first show that
	$\int_{\mathbb{R}^d} |g_n(x) - g_0(x) |\,\textup{d}x \xrightarrow[n\to\infty]{P_{\lambda_0,\zeta_0}} 0$. Denote the $\epsilon$ chosen from Lemma \ref{lemma:LAN} as $\epsilon_0$.
	Since $\pi_0$ is continuous at $\lambda_0$, we choose sufficiently small $\epsilon\in (0,\epsilon_0 / 2)$ such that $\pi_0(\lambda) \leq 2\pi_0(\lambda_0) $ for all $\lambda\in B_{2\epsilon}(\lambda_0)$. Let $\delta$ be the number from the condition.
	
	\citet[Corollary 1]{murphy2000profile} shows that $\sqrt{n}\|\lambda_n - \lambda_0\|$ is bounded in probability and for all $\lambda\in B_{\epsilon_0}(\lambda_0)$ and large enough $n$ such that $\lambda_n \in B_{\epsilon_0}(\lambda_0)$,
	\be
	\hat{l}_{n}(\lambda)- \hat{l}_{n}(\lambda_n) = -\frac{1}{2}  (\lambda - \lambda_n)^\top H_0(\lambda-\lambda_n)
	+ o_{P_{\lambda_0,\zeta_0}}(1)\bigl\{\bigl(\|\lambda-\lambda_n\| + n^{-1/2} \bigr)^2\bigr\}.
	\ee
	Letting $\lambda = \lambda_n + x/\sqrt{n}$ with $x\in B_{\epsilon\sqrt{n}}(0)$ and large enough $n$ such that $\lambda_n \in B_{\epsilon_0 / 2}(0)$, we have
	\(
	n\bigl\{\hat{l}_n(\lambda_n + x/\sqrt{n}) - \hat{l}_n(\lambda_n)\bigr\} = -\frac{1}{2} x^\top H_0 x + o_{P_{\lambda_0,\zeta_0}}(1) (\|x\| + 1)^2.
	\)
	Combining with $\pi_0$ is continuous at $\lambda_0$ and $\lambda_n+x/\sqrt{n}\to \lambda_0$, we have $g_n(x)\to g_0(x)$ pointwise with probability converge to $1$.
	Consider $n>1/\epsilon ^2$ sufficiently large such that the term $o_{P_{\lambda_0,\zeta_0}}(1) < \alpha / 4$ where $\alpha$ is less than the smallest eigenvalue of $H_0$. We denote $A_0 = H_0 - \alpha I$, and define
	\(
	h_n(x) = \begin{cases}
		\exp( - x^\top A_0 x /2 + \alpha / 2) 2\pi_0(\lambda_0) & \text{if } \|x\| < \epsilon\sqrt{n},\\
		\exp(-n\delta / 2)\pi_0(\lambda_n+x/\sqrt{n}) & \text{if } \|x\| \geq  \epsilon\sqrt{n}.
	\end{cases}
	\)
	When $\|x\| <  \epsilon\sqrt{n}$, for $n$ large enough we have $\|(\lambda_n+x/\sqrt{n}) - \lambda_0\| < \|\lambda_n-\lambda_0\| + \epsilon < 2\epsilon$. By the choice of $\epsilon$, we have $\pi_0(\lambda_n+x/\sqrt{n}) \leq 2\pi_0(\lambda_0)$. Since $(\|x\| + 1)^2 \leq 2\|x\|^2 + 2$, we have $o_{P_{\lambda_0,\zeta_0}}(1) (\|x\| + 1)^2\leq \alpha (\|x\|^2 + 1) / 2 = \alpha x^\top x / 2 + \alpha / 2 $. Hence $g_n(x) \leq h_n(x)$ with probability converge to $1$ for $n$ sufficiently large, combining with the condition in theorem when $\|x\| \geq  \epsilon\sqrt{n}$.
	
	Also, $h_n(x) \to h_0(x) = \exp\{ - x^\top A_0 x /2 + \alpha / 2\} 2\pi_0(\lambda_0)$ pointwise. Now,
	\(
	\int_{\mathbb{R}^d} h_n(x)\,\textup{d}x  = \int_{\|x\| < \epsilon\sqrt{n}} \exp( - x^\top A_0 x /2) e^{\alpha / 2} 2\pi_0(\lambda_0) \,\textup{d} x + \int_{\|x\| \geq  \epsilon\sqrt{n}} e^{-n\delta / 2}\pi_0(\lambda_n+x/\sqrt{n}) \,\textup{d}x.
	\)
	The second term is less than $\int_{\mathbb{R}^d} e^{-n\delta / 2}\pi_0(\lambda_n+x/\sqrt{n}) \,\textup{d}x = e^{-n\delta / 2} \int_{\mathbb{R}^d}\pi_0(\lambda) n^{d/2} \,\textup{d}\lambda = e^{-n\delta / 2} n^{d/2} \to 0$, while the first term monotonically converges to $\int_{\mathbb{R}^d} h_0(x)\,\textup{d}x$. Since $g_n,g_0,h_n,h_0$ are integrable, by the generalized dominated convergence theorem (the version for convergence in probability), we have $\int_{\mathbb{R}^d} |g_n(x) - g_0(x) |\,\textup{d}x \xrightarrow[n\to\infty]{P_{\lambda_0,\zeta_0}} 0$ and $\int_{\mathbb{R}^d} g_n(x) \,\textup{d}x \xrightarrow[n\to\infty]{P_{\lambda_0,\zeta_0}} \int_{\mathbb{R}^d} g_0(x) \,\textup{d}x$.
	
	Let $a_n = 1/\int_{\mathbb{R}^d} g_n(x)\,\textup{d}x$ and $a_0 = 1/\int_{\mathbb{R}^d} g_0(x)\,\textup{d}x $. Then $a_n\to a_0$ in $P_{\lambda_0,\zeta_0}$-probability, and thus
	\(
	&\int_{\mathbb{R}^d} \bigg|q_n(x) - \mathcal{N}\bigl(x \mid 0, H_{0}^{-1}\bigr) \bigg|\,\textup{d}x
	= \int_{\mathbb{R}^d} |a_n g_n(x) - a_0 g_0(x) |\,\textup{d}x \\
	& \leq \int_{\mathbb{R}^d} |a_n g_n(x) - a_n g_0(x)|\,\textup{d}x + \int_{\mathbb{R}^d} |a_n g_0(x) - a_0 g_0(x) |\,\textup{d}x\\
	&  \leq |a_n| \int_{\mathbb{R}^d} |g_n(x)-g_0(x)|\,\textup{d}x + |a_n-a_0| \int_{\mathbb{R}^d} |g_0(x)|\,\textup{d}x \xrightarrow[n\to\infty]{P_{\lambda_0,\zeta_0}} 0.
	\)
	This proves $\text{d}_{\text{TV}}\bigl\{q_n,\mathcal{N}\bigl(0,H_0^{-1}\bigr)\bigr\} \xrightarrow[n\to\infty]{P_{\lambda_0,\zeta_0}} 0$, and $\int_{B_{\epsilon}(\lambda_{0})} \pi_{n}(\lambda)\, \textup{d} \lambda \xrightarrow[n\to\infty]{P_{\lambda_0,\zeta_0}} 1$ follows from \citep[Lemma 28]{miller2021asymptotic}.
\end{proof}

\section{An Illustration of Least Favorable Submodel}

The Hilbert space $\widetilde{\mathcal{H}}$ highly depends on the parameter space $\mathcal{H}$. To illustrate how this works, we show an example from the Cox regression model without censoring.

\begin{example}[Cox regression model without censoring] Consider the density function of the survival time $(y_1,\dots,y_n)$ where $y_i\in \mathbb{R}_+:\mathbb{R}\cap [0,\infty]$ with covariates $(x_1,\dots,x_n)$ where $x_i\in \mathbb{R}$:
	$$
	L(y_{1:n},\zeta;\lambda) = \prod_{i=1}^n \exp(\lambda x_i) \zeta(y_i) \exp\{-\exp(\lambda x_i) Z(y_i)\},
	$$
	where the parameter $\lambda\in\mathbb{R}$ and $Z(y)=\int_{0}^y \zeta(y) \, \textup{d}y$. The latent variable  $\zeta$ is a ``hazard function", which is a non-negative integrable function on $\mathbb{R}_+$, and belongs to the Hilbert space $\mathcal{H}=L^1(\mathbb{R}_+)$. The ``cumulative hazard function" $Z$ is a non-negative and non-decreasing function on $\mathbb{R}_+$. In regard to the model, we have
	$$
	l_n(\lambda, \zeta)= \sum_{i=1}^n \{ \lambda x_i + \log \zeta(y_i) - \exp(\lambda x_i) Z(y_i)\} / n,
	$$
	and the $\lambda$-score function is $\dot{l}_n(\lambda,\zeta) = \sum_{i=1}^n \{x_i - x_i \exp(\lambda x_i) Z(y_i)\} / n$.
	
	Let $\widetilde{H} = L^2(\mathbb{R}_+)$. Given a fixed function $\zeta_0$ and a bounded function $\delta\in \widetilde{H}$, we can define a path $\{\zeta_\gamma^\delta \in \mathcal H\}_{\gamma\in \mathbb{R}}$ by $\zeta_\gamma^\delta(y) = \{1+(\gamma - \lambda_0)\delta(y)\} \zeta_0(y)$ for all $y\in \mathbb{R}_+$. It satisfies that $\zeta^\delta_\gamma \to \zeta_0$ in $\mathcal{H}$ when $\gamma\to \lambda_0$. Also, we define $Z_\gamma^\delta(y) = \int_0^y \{1+(\gamma-\lambda_0)\delta(y)\} \zeta_0(y)\, \textup{d}y$ correspondingly.
	Now, plugging $\zeta^\delta_\gamma$ into $l_n(\lambda_0, \zeta)$ as $\zeta$ and differentiating it at $\gamma=\lambda_0$, we get the $\zeta$-score function at $\zeta=\zeta_0$ in the direction of $\delta$ by
	$$
	A^n_{\lambda_0, \zeta_0} \delta = \sum_{i=1}^n \{\delta(y_i) - \exp(\lambda_0 x_i) \int_0^{y_i} \delta(y) \zeta_0(y)\,\textup{d}y \} / n.
	$$
\end{example}

\begin{figure}[H]
    \centering
    \begin{subfigure}[t]{0.4\textwidth}
        \centering
        \includegraphics[width=1\linewidth]{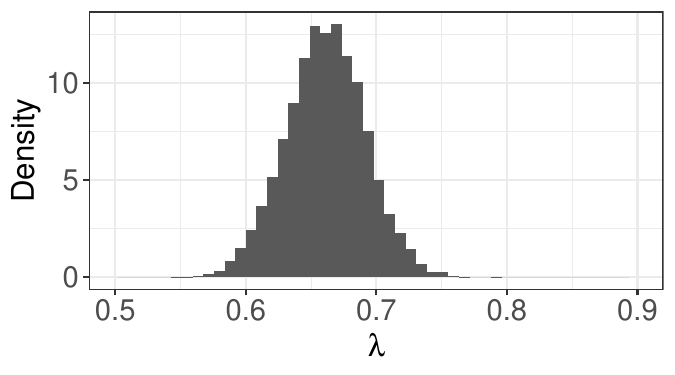}
        \caption{Posterior distribution of $\lambda$  from the full Bayesian model.} 
    \end{subfigure}
    \begin{subfigure}[t]{0.4\textwidth}
        \centering
        \includegraphics[width=1\linewidth]{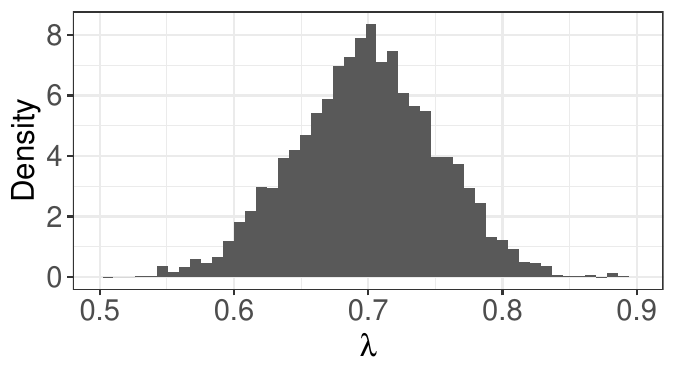}
        \caption{Posterior distribution of $\lambda$ from the bridged posterior.}
    \end{subfigure}
    \caption{The posterior densities of the parameter $\lambda$ in Cox regression model under sample size 500.
    \label{fig:cox_posterior}}
\end{figure}

We conduct a numerical experiment for the Cox regression model. We generate the data with sample size $n = 500$. We sample the covariate $(x_1,\dots,x_n)$ independently from a standard normal distribution. We create a baseline hazard function $\zeta_0$ with piecewise constant over 5 intervals, where hazard rates are assigned to each of these intervals. We set the ground-truth parameter of interest $\lambda_0$ as $0.8$. Then we sample the survival time $(y_1,\dots,y_n)$ from the hazard function $\exp(\lambda_0 x_i) \zeta_0(y_i)$. For the bridged posterior based on profile likelihood, $g(\zeta,y;\lambda) = -L(y,\zeta;\lambda)$, we set N$(0, 5^2)$ prior on $\lambda$. For the canonical integrated posterior model, we set the same prior on $\lambda$ and  Gamma$(1,1)$ priors on hazard rates in each interval. We run the MCMC samplers for $10000$ iterations and discard the first $2000$ as burn-ins. Figure \ref{fig:cox_posterior} shows that the posterior distributions for $\lambda$ under the two models are similar to each other.

\section{Comparison with Existing BvM Results on Semi-parametric Models}\label{sec:compare_bvm}

There is a rich theory literature on BvM results on semi-parametric models. Naturally, it is of interest to compare our results with them, contextualizing and clarifying our contribution. The existing results can roughly be divided into two categories. The first is similar to our focused setting where the posterior is obtained under a profile likelihood. \cite{lee2005profile} showed that $\mathbb{E}_{\lambda\sim \Pi(\lambda\mid y)} g\{\sqrt{n}(\lambda-\lambda_n)\}$ converges to $\mathbb{E}_{u\sim \mathcal{N}(0,\tilde{I}_0^{-1})} g(u)$ in probability, assuming a Taylor expansion form and for iid data. Their condition is similar to \eqref{eq:lan}, and on the other hand, they do not give the result of the posterior density converging to normal density in total variation.
\cite{cheng2008higher} showed a BvM result for the posterior induced from profile likelihood for iid probability model, under the assumption that the third derivatives exist. Compared to their result, ours is general in the sense that it is applicable to non-iid data and under potential non-differentiability.

The second category of BvM results relate to canonical Bayesian methodology involving integrated posterior $\Pi(\lambda\mid y)= \int\Pi(\lambda,\textup{d} \zeta\mid y)$ over a non-deterministic $\zeta$. \cite{bickel2012semiparametric} proved a BvM result for marginal posterior distribution of $\lambda$ using the LAN property for the marginal likelihood of $\lambda$, which has similar form with \eqref{eq:lan}. On the other hand, they additionally assume that the marginal posterior probability of $\lambda$ inside the neighborhood $B_{M_n/\sqrt{n}}(\lambda_0)$ converges to $1$ for every $M_n\to\infty$. This condition is similar to but arguably stronger than the last condition in Theorem \ref{thm:BvM_semi}. \cite{castillo2015bernstein} proved a BvM result on a functional of the parameters, under an essentially necessary no-bias condition which is related to both the likelihood and the prior specification of ($\lambda, \zeta$).

In \cite{bickel2012semiparametric}, to get the LAN property for the marginal likelihood of $\lambda$, they first assume that a neighborhood of $z$ has enough prior mass, and the parameter space of $z$ has bounded Hellinger metric entropy (covering number). They assume that for $z$ outside a neighborhood of the fixed $z_0$, the Hellinger distances between the likelihood at $\lambda_0+h_n/\sqrt{n}$ and the one at $\lambda_0$ are uniformly (with respect to $z$) infinitesimal for every bounded $h_n$. Finally,
assuming that the least favorable submodel exists, that is a submodel $l_n(\lambda,z_\lambda)$ with parameters $(\lambda,z_\lambda)$ satisfying $\mathscr{Q}\dot{l}_n(\lambda,z_\lambda) = \dfrac{\partial l_n(\lambda,z_\lambda)}{\partial \lambda}$ for all $\lambda$ in a neighborhood of $\lambda_0$, the conditional posterior distribution of $z$ can be shown to concentrate around the parameter of the least favorable submodel $z_\lambda$, that is the probability that the Hellinger distance between $z$ and $z_\lambda$ is greater than a positive number converges to zero. This leads to the marginal LAN property assuming that the full likelihood has LAN property in the direction of $\lambda$ when the $z$ is perturbed around the least favorable submodel, that is $z=z_\lambda+\zeta$ for all $\zeta$ in a neighborhood of $0$.

In \cite{castillo2015bernstein}, let $h$ be the least favorable direction satisfying \\ $\mathscr{P} \dot{l}_n(\lambda_{0}, z_{0})  =A_{\lambda_{0}, z_{0}} h.$
Let $\lambda_t=\lambda-t\tilde{I}_0^{-1}/\sqrt{n}, $ and $z_t=z+th\tilde{I}_0^{-1}/\sqrt{n}$. Assume over the direction of $h$, the log full likelihood has the LAN property with a remainder term, and the difference between the remainder terms at $(\lambda,z)$ and at $(\lambda_t,z_t)$ converges to zero. Further, suppose that the ratio between the integral of the likelihood under the prior at $(\lambda_t,z_t)$ and the one at $(\lambda,z)$ converge to $1$ for all $(\lambda,z)$. They proved the BvM theorem where the mean is $\lambda_0$ plus the first order term of the LAN expansion.

\section{Data Augmentation for Latent Normal Model}\label{sec:data_aug_latent}
For the latent normal model with binary observations, we follow \cite{polson2013bayesian} and use the following data augmentation:
\(
L(\lambda,\zeta,\eta ; y) \propto \exp\biggl\{ -\frac{1}{2} \zeta^\top Q^{-1} (\lambda;x) \zeta \biggr\} \prod_{i=1}^n \exp\{ (y_i-1/2) \zeta_i \} \exp\{ -\eta_i(\zeta_i)^2/2 \} \text{PG}(\eta_i; 1,0) \pi_0(\lambda)  ,\)
where $\textup{PG}(\cdot; 1,0)$ is the density of P\'olya-Gamma$(1,0)$ distribution. This leads to closed-form update of $\zeta$ from a normal full conditional distribution.

\section{Additional Simulation Results}

\subsection{Results of Posterior Approximation for Latent Normal Model}\label{sec:approx_latent}
\begin{figure}[H]
    \centering
        \begin{subfigure}[t]{0.35\textwidth}
        \centering
        \includegraphics[width=1\linewidth]{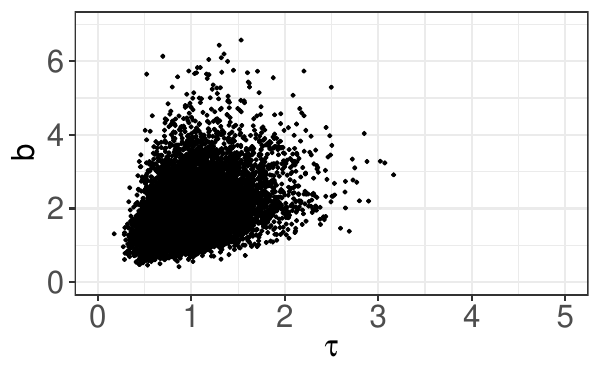}
        \caption{Posterior distribution of $(b,\tau)$ from INLA.}
    \end{subfigure}
    \begin{subfigure}[t]{0.35\textwidth}
        \centering
        \includegraphics[width=1\linewidth]{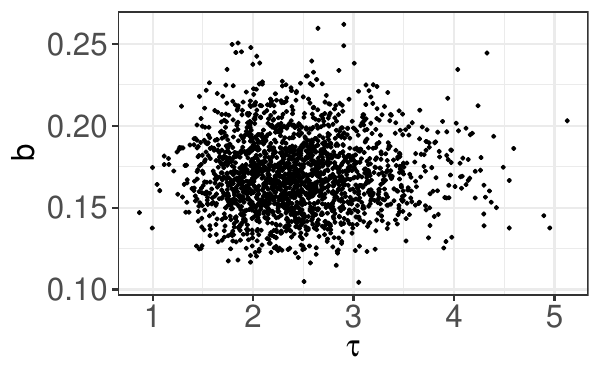}
        \caption{Posterior distribution of $(b,\tau)$  from variational inference.} 
    \end{subfigure}
    \caption{The posterior distributions of the covariance kernel parameters from the latent normal model (with sample size 1000) using integrated nested Laplace approximations (Panel a) and variational inference (Panel b) .
    \label{fig:latent_gau_posterior_vi}}
\end{figure}

We conduct simulated experiments to show results of posterior approximation for the  latent normal model. We use two algorithms: the integrated nested Laplace approximations (INLA) and the variational inference with mean field approximation. For the INLA, we use \texttt{inla} function with default option from the R package \texttt{INLA}. For the variational inference, we use normal distribution with a diagonal covariance matrix as the variational distribution, and \texttt{pyro} package in python. 
Figure \ref{fig:latent_gau_posterior_vi} shows that the INLA produces an distribution estimate similar to the exact posterior. On the other hand, the variational inference severely underestimates the value of $b$.

\subsection{Simulation for Comparing the Posterior Variances from Latent Normal Model and Latent Quadratic Exponential Model}\label{sec:sim_latent_normal_var}

We use a simulation experiment of latent quadratic exponential model to show that the profile likelihood-based bridged model has an asymptotic posterior variance for the parameter $\lambda$ that is equal to that of the Bayesian model based on the full likelihood. 

We generate random locations $x_1,\dots,x_S \sim \text{Uniform}(-6, 6)$ where $S\in \{50,200,500,1000\}$ is the sample size, and ground-truth means from $6$ latent curves $\tilde{z}_{ji} = f_j(x_i)$ where $j=1,\dots,6$. At each $x_i$ for each curve, we generate a binary $y_{ji} \sim \text{Bernoulli}(1/\{1+\exp(-\tilde{z}_{ji})\})$.

For each group of data $y_{j1},\dots,y_{jS}$ from the $j$-th curve, we fit both the latent quadratic exponential model and the latent normal model. For both model, we assign a half-normal $\text{N}_{+}(0, 1)$ prior on $\tau$ and Inverse-Gamma$(2, 5)$ prior on $b$. We use the same algorithm that is used in \cref{subsec:simulation_latent}. Since the latent quadratic exponential model enjoys much better mixing performance compared to the latent normal model, for the former model, we run the MCMC algorithm for $5000$ iterations, discard the first $2000$ as burn-ins and the samples are thinned at $10$, while for the latter one, we run the MCMC algorithm for $13000$ iterations, discard the first $4000$ as burn-ins,  and the samples are thinned at $30$.

\subsection{Flow network problem with uncertainty on some capacity values}\label{sec:flow}

Consider a directed network $G=(V,E)$ with $V$ the set of nodes, and $E$ the set of uni-directed edges $(i\to j)$, each associated with a capacity value $c_{ij}>0$. We have observation of flow $y_{ij}\in [0,c_{ij}]$ on $(i\to j)\in E$, and $y_{ij}=0$ if $(i\to j)\not \in E$. Geography studies often impose a mechanistic model where the flow network is operating close to its maximum capacity when the network is congested, associated with the following linear program:
\(
\quad & g(\zeta;\lambda)=\sum_{j:\,(s\to j) \in E} \zeta_{sj} \\
\text{subject to} \quad & \sum_{j:\,(i\to j) \in E} \zeta_{ij} \, \, - \sum_{k:\,(k\to i) \in E} \zeta_{ji} = 0, \quad \forall i \in V \setminus \{s, t\},  \\
&0 \le \zeta_{ij} \le \lambda_{ij}, \quad \forall (i\to j) \in E_*,\\
& 0 \le \zeta_{ij} \le c_{ij}, \quad \forall (i\to j) \in E \setminus E_*,  
\)
where $s$ is the source node where there is a positive total flow entering the node, and $t$ is the sink node where there is a negative total flow corresponding to leaving the network. The equality constraints above correspond to flow conservation at each node that is neither source or sink. In practice, we often have a subset of edges $E_*$ where the capacity values $c_{i,j}$ are not constant but contain great uncertainty (such as roads that are prone to accidents), motivating for a statistical model where those $c_{ij}$ are replaced by parameters $\lambda_{ij}$. Taking consideration of the measurement error in $y_{ij}$, we have the following bridged posterior:
$$
L(y,z;\lambda) \pi_0(\lambda)  \propto  \Biggl[ \prod_{k=1}^n  (\sigma^2)^{-|E|/2}\exp\biggl\{ -\frac{ \sum_{(i\to j)\in E}(y^k_{ij}-z_{ij})^2}{2\sigma^2}
\biggr\} \Biggr] \exp( - \rho\sum_{(i\to j)\in E_*}\lambda_{ij})
$$
where $z={\arg\min}_\zeta g(\zeta;\lambda)$, and we assign an exponential prior for each $\lambda_{ij}$.

\begin{figure}[H]
    \centering
    \includegraphics[width=.19\textwidth]{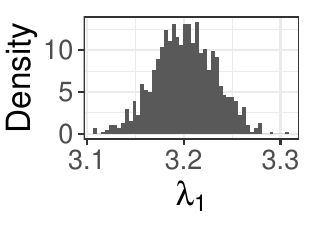}
    \includegraphics[width=.19\textwidth]{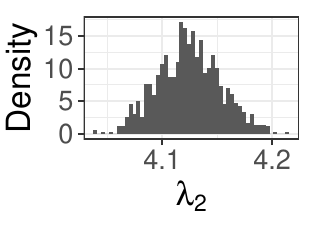}
    \includegraphics[width=.19\textwidth]{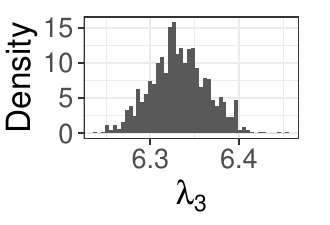}
    \includegraphics[width=.19\textwidth]{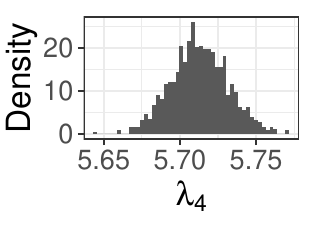}
    \includegraphics[width=.19\textwidth]{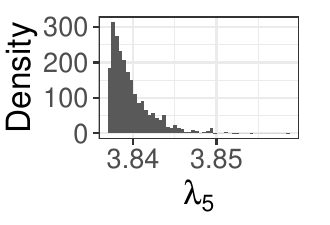}
    \caption{Flow value posteriors of $\lambda_{ij}$ for each network edge in $E_*$ from the bridged posterior; vertical lines depict ground-truth values $c_{ij}$.}
    \label{fig:flow-net-lambda-hist}
\end{figure}

\begin{figure}[H]
    \centering
    \includegraphics[width=.45\textwidth]{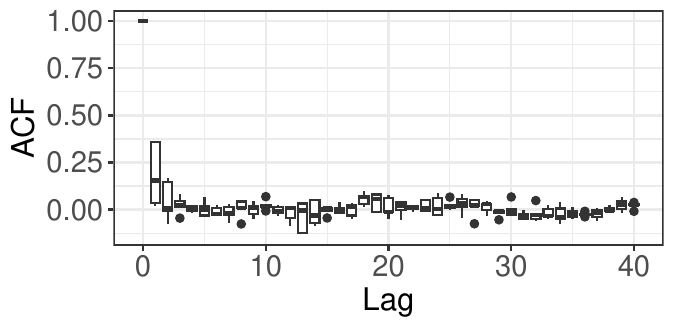}
    \includegraphics[width=.45\textwidth]{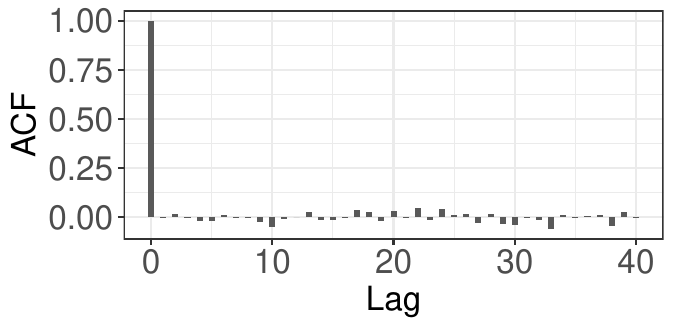}
    \caption{The autocorrelation functions for posterior Markov chains of $\lambda_{ij}$'s (by boxplots, left) and $\sigma^2$ (right).}
    \label{fig:flow-net-acf}
\end{figure}

We use \texttt{networkx} to generate a directed network with $40$ nodes and $371$ edges, and make sure there is one source and one sink. We generate the capacities $c_{ij}$ independently from $\text{Uniform}(2,10)$. The ground-truth maximum flows $z^0_{ij}$ are computed using built-in function from \texttt{networkx}.  We select $5$ edges to form the set $E_*$; the corresponding $\lambda_{ij}$'s are the low-dimensional parameters of interest. To simulate the observations, we generate $n=500$ samples $y_{ij}^k \sim \text{N}(z^0_{ij}, 1.0)$ for each edge $(i\to j)\in E$. For the prior, we set $\rho=0.2$, and we use an Inverse-Gamma$(2,5)$ prior on $\sigma_2^2$. We run the random walk Metropolis--Hastings algorithm for $10000$ iterations, discarding the first $2000$ as burn-in and thinning at $10$.
Figure \ref{fig:flow-net-acf} show the good mixing performance of the posterior sampling algorithm, which takes $0.045$ seconds per iteration on a 4.0 GHz processor.  Figure \ref{fig:flow-net-lambda-hist} shows the posterior densities of $\lambda_{ij}$'s for $(i\to j)\in E_*$.

\section{ADMM for Optimization Problem in Data Application} \label{sec:admm}
To solve the optimization problem
\(
\min_{\zeta} \frac{1}{2}\|\mathcal L-\zeta\|^2 + \tilde\lambda \|\zeta\|_* \quad \text{subject to }  \zeta\in \mathbb{R}^{n\times n},  \zeta_{i,i}=\sum_{j:j\neq i}\zeta_{i,j} , \zeta_{i,j}=\zeta_{j,i} \leq 0 \text{ for } i \neq j,
\)
we use ADMM under constraints and log-barrier:
\(
& \min_{\zeta, Z} \frac{1}{2}\|{\mathcal L}-\zeta\|^2 +  \rho\sum_{(i,j):i \neq j}\{-\log (-\zeta_{i,j})\}+ \tilde\lambda \|Z\|_* + \dfrac{\eta}{2}\|\zeta-Z+W\|^2\\
& \text{ subject to } \zeta_{i,i}=-\sum_{j:j\neq i}\zeta_{i,j} , \zeta_{i,j}=\zeta_{j,i} \leq 0 \text{ for } i \neq j,
\)
where $W=W^T$ is the Lagrangian multiplier. The ADMM algorithm iterates the following steps:
\begin{enumerate}[leftmargin=1cm]
	\item Constrained gradient descent for $\zeta$: set $\zeta$ to be
	\(
	& {\arg\min}_\zeta \frac{1}{2}\|{\mathcal L}-\zeta\|^2 +  \rho\sum_{(i,j):i \neq j}\{-\log (-\zeta_{i,j})\}+ \frac{\eta}{2}\|\zeta-Z+W\|^2 \\
	& \text{subject to }  \zeta_{i,i}=-\sum_{j:j\neq i}\zeta_{i,j} , \zeta_{i,j}=\zeta_{j,i} \text{ for } i \neq j.
	\)
	The constraints are easy to satisfy, by restricting the free parameters to $\{\zeta_{i,j}\}_{i>j}$, and setting $\zeta_{i,i}= -\sum_{j<i} \zeta_{i,j}   -\sum_{j>i} \zeta_{j,i}$.
	\item Minimizing over $Z$: set $Z= S_{\tilde\lambda/\eta}(\zeta+W)$
	where $S_{\tilde\lambda/\eta}(X)
	= \sum_{i=1}^n (\sigma_i - \tilde\lambda/ \eta)_+ u_iv_i^T$, and $X=U\text{diag}(\sigma_i)V$ is the singular value decomposition. The solution of this step satisfies the conditions of symmetry and the rows add to zero.
	\item Updating $W$: set $W$ to be $\zeta-Z+W $.
\end{enumerate}

\end{document}